\documentclass{egpubl}
\usepackage{eg2024}
 
\ConferencePaper       
\CGFccby

\usepackage[T1]{fontenc}
\usepackage{dfadobe}  

\usepackage{cite}
\biberVersion
\BibtexOrBiblatex
\electronicVersion
\PrintedOrElectronic
\ifpdf \usepackage[pdftex]{graphicx} \pdfcompresslevel=9
\else \usepackage[dvips]{graphicx} \fi

\usepackage{egweblnk} 
\usepackage{amssymb}
\usepackage{amsmath}

\usepackage[ruled]{algorithm2e} 
\usepackage{booktabs} 
\usepackage{dsfont}

\title{CharacterMixer: Rig-Aware Interpolation of 3D Characters}

\author[X. Zhan, R. Fu, D. Ritchie]
{
\parbox{\textwidth}
{
\centering
X. Zhan$^1$\orcid{0000-0003-1375-180X}
\hspace{0.1in}
R. Fu$^{1}$\orcid{0000-0002-0115-0831} 
\hspace{0.1in}
D. Ritchie$^{1}$\orcid{0000-0002-8253-0069}
}
\\
{\parbox{\textwidth}
{\centering
$^1$Brown University, USA\\
}
}
}
 
\newcommand{\methodname}{CharacterMixer}

\begin{document}

\teaser{
\vspace{-0.25in}
 \includegraphics[width=0.95 \linewidth]{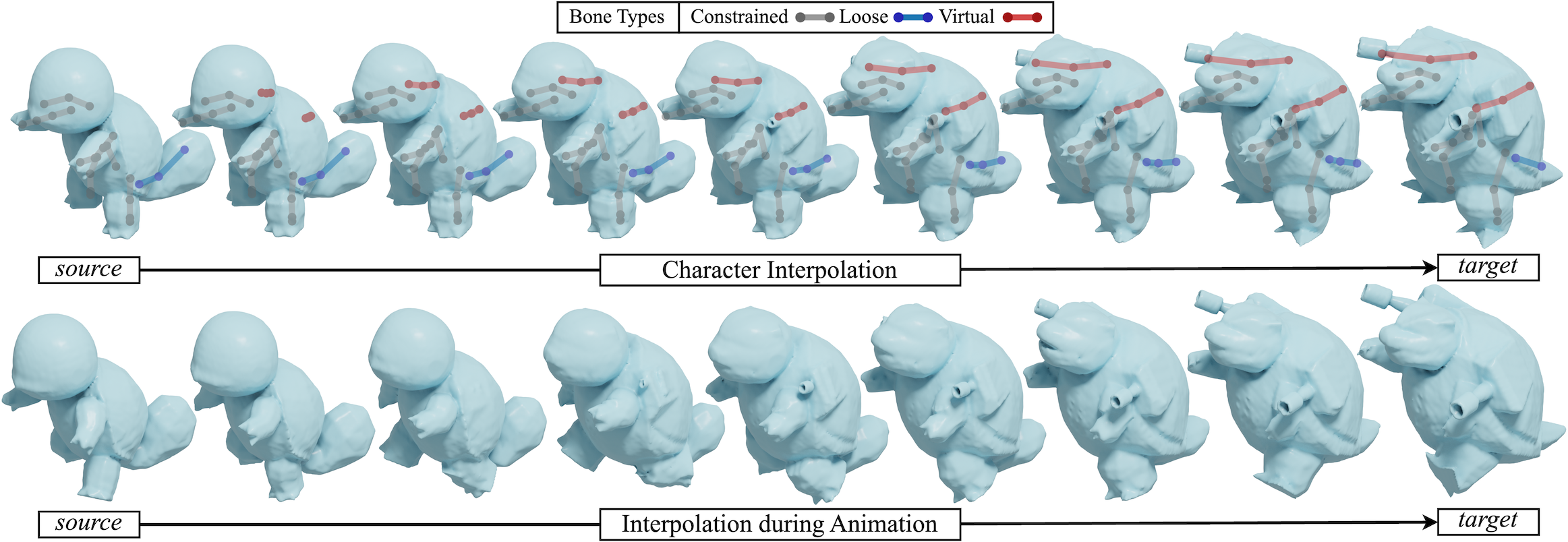}
 \centering
  \caption{
\methodname{} enables interpolation between two 3D models that have different surface mesh topologies and rig skeletons.
It preserves a posable rig throughout interpolation.
In the top row, we show an example of interpolating source to target with a fixed pose.
\methodname{} constructs three types of bones to handle different skeleton topologies: 1-to-1 matched bones are Constrained, 1-to-many matches are Loose (tail), and 1-to-void matches are Virtual (shoulder-mounted cannons).
The bottom row shows the same interpolation, but where the character's pose changes continuously during a run cycle (posed rigs are omitted for a clear visual of the geometry).
}
\label{fig:teaser}
}

\maketitle

\begin{abstract}
We present CharacterMixer, a system for blending two rigged 3D characters with different mesh and skeleton topologies while maintaining a rig throughout interpolation.
CharacterMixer also enables interpolation during motion for such characters, a novel feature.
Interpolation is an important shape editing operation, but prior methods have limitations when applied to rigged characters: they either ignore the rig (making interpolated characters no longer posable) or use a fixed rig and mesh topology.
To handle different mesh topologies, CharacterMixer uses a signed distance field (SDF) representation of character shapes, with one SDF per bone.
To handle different skeleton topologies, it computes a hierarchical correspondence between source and target character skeletons and interpolates the SDFs of corresponding bones.
This correspondence also allows the creation of a single ``unified skeleton'' for posing and animating interpolated characters.
We show that CharacterMixer produces qualitatively better interpolation results than two state-of-the-art methods while preserving a rig throughout interpolation.
Project page: \href{https://seanxzhan.github.io/projects/CharacterMixer.html}{https://seanxzhan.github.io/projects/CharacterMixer}.

\ccsdesc[500]{Computing methodologies~Shape analysis}
\ccsdesc[500]{Computing methodologies~Shape modeling}

\printccsdesc   
\end{abstract}  

\section{Introduction}

Interpolation is a fundamental operation in 3D shape modeling and editing.
Producing smooth blends between shapes can be used to create animations~\cite{BlenderShapeKeys}, to ``fill in gaps'' between shapes in a collection~\cite{glass2022sanjeev}, or to create new hybrid shapes~\cite{StructuralBlending}.
One of the most common types of 3D shape is a \emph{3D character}: an articulated body that is animated in some film, game, or other 3D graphics experience.
Interpolation between 3D characters can be used for pose matching~\cite{NeuroMorph} or for creating a range of blended characters from a smaller set of hand-modeled ones (e.g. for creating crowds of background characters)~\cite{PixarBackgroundChars}.

When the shapes to be interpolated are 3D characters, the \emph{rigs}, or the articulated skeletons that allow characters to be animated, for those characters should be taken into account, which complicates the problem.
In practice, most systems which can interpolate between rigged characters are based on parametric models which can produce variations of the character's body shape but always keep the same surface mesh and rig topologies, limiting the range of characters that can be interpolated~\cite{SMPL:2015}.
Methods exist for interpolating between different 3D shapes, but when applied to rigged characters, they ignore the rigs, leading to intermediate shapes that are no longer rigged and thus not directly posable~\cite{NeuroMorph}. 

In this paper, we present \methodname{}, the first system for interpolating between two rigged characters with different mesh and skeleton topologies, such that a rig is preserved.
With the preserved rig, not only does \methodname{} allow interpolated characters to be posable, but it can also generate animation sequences where interpolation happens at the same time (Fig.~\ref{fig:teaser}).
Handling different mesh and skeleton topologies is crucial for interpolation tasks involving characters not created by the same artist, or when animators do not control characters' sources, such as uploaded assets in the online gaming community. 
To handle characters with different mesh topologies, \methodname{} uses signed distance field (SDF) representations of the source and target geometries.
To make the system rig-aware, a character is represented as a union of SDFs, one per each bone of the rig.
Mesh-based methods such as NeuroMorph~\cite{NeuroMorph} are unable to interpolate the identities of characters; they deform the source mesh to match the shape of the target, keeping the source topology unchanged.
In contrast, the SDF representation allows our method to interpolate geometry and produce intermediate characters of different identities (Fig.~\ref{fig:teaser}).
To interpolate between two rigged characters with different skeletal topologies, \methodname{} computes a hierarchical correspondence between two skeletons.
This correspondence allows it to create a single ``unified skeleton'' whose pose can drive the pose of both the source and target characters.
Given the unified skeleton, \methodname{} interpolates between the two characters' geometries by linear interpolation of the SDFs of corresponding bones.

We evaluate \methodname{} by comparing to a state-of-the-art optimal transport approach for shape interpolation~\cite{ConvWasser} and a mesh-based data-driven method for shape correspondence and interpolation~\cite{NeuroMorph}, showing that \methodname{} generates intermediate shapes with higher visual fidelity while also maintaining a posable rig.
In summary, our contributions are:
\begin{itemize}
    \item A method for computing hierarchical correspondence between two skeletons and producing unified intermediate skeletons
    \item A technique for posing and animating interpolated characters using the unified skeletons
    \item An interpolation approach for blending between two characters' geometries while preserving a rig
\end{itemize}

\section{Related Works}

\textbf{Shape Interpolation.}
There is a significant body of prior work on shape interpolation and blending.
One family of work uses optimal transport, treating the source and target shapes as probability distributions and finding a transformation of the source to the target that moves as little probability mass as possible~\cite{ConvWasser,DebiasedSinkhorn,DiffAqua}.
Another work interpolates the interiors of shapes in an as-rigid-as-possible manner, restricting local volumes to be least-distorting~\cite{ARAPInterp}.
There are also data-driven approaches, interpolating from a source shape to a target shape by finding a path through a large collection of related shapes~\cite{PartDataInterp,MorphingEditing,RealisticShapeMorphing} 
or using the structures of manufactured shapes~\cite{DSGNET, SDMNET}.
Most recently, several works train neural networks to produce deformations from a source to target shape~\cite{NeuroMorph,Yifan:NeuralCage:2020,jiang2020shapeflow}.
Deep generative models can also be viewed as interpolators, as their latent spaces allow interpolation between shapes in the generator's output domain~\cite{Achlioptas2017LearningRA,SPGAN,PointFlow,chen2018implicit_decoder,zheng2022sdfstylegan}.
These methods are all oblivious to character rigs and thus intermediate interpolated shapes would not be posable, making it impossible to interpolate throughout an animation.

To the best of our knowledge, no prior work focuses on rig-aware character interpolation.
Parametric body models, such as SMPL~\cite{SMPL:2015}, support interpolation between body shapes with the same rig; these shapes all have the same mesh and skeleton topology.
Our method supports interpolation between characters with different mesh and skeleton topologies.

\textbf{Automated Character Rigging.}
Our system interpolates between 3D characters such that the intermediate characters are still animatable.
One could instead use a rig-oblivious shape interpolation method and then attempt to automatically compute a rig for the new intermediate shape.
Several automated rigging methods exist: some are restricted to characters created via a specialized sketch-based modeling interface~\cite{RigMesh,MonsterMash}, whereas others can take arbitrary shapes as input and produce a skeleton~\cite{AnimSkelVolNet}, potentially with skinning weights~\cite{RigNet}.
These methods can sometimes fail to predict usable rigs, and they would produce different rigs for each step in an interpolation sequence.
Our method produces a single rig that can animate all intermediate characters over an interpolation sequence.
Alternatively, one may opt not to use a skeleton to pose an intermediate character.
This would require posing one of the two input characters and using techniques such as ~\cite{SurfaceBasedMotionRetargeting, RealTimeMotionRetargeting, ContactAwareRetargeting, MoCaNet} to transfer the pose to the other character before blending the two characters. 
However, this offers no direct rigging control over the intermediate character.

\textbf{Tree Correspondence.}
\methodname{} computes a hierarchical correspondence between two rig skeletons.
This is related to prior work on computing hierarchical correspondences between 3D shapes~\cite{DeformShapeCorres}.
One work in this space uses these correspondences to ``interpolate'' between shapes~\cite{StructuralBlending}, though they focus on manufactured objects and produce transitions that involve discrete structural switches; we instead focus on continuous blends.

\textbf{Part-based character blending.}
One way to produce transitions or blends between two 3D shapes is by gradually swapping their constituent parts.
Modeling-by-assembly could be used to do this, albeit with considerable user interaction~\cite{ModelingByExample,Shuffler}.
Some prior work can produce such transitions automatically~\cite{PartBasedRecombination}.
There also exists a system which can mix and match parts from rigged character models such that the resulting chimera is also rigged~\cite{SkinMixer}.
Our goal is to produce a qualitatively different kind of interpolation between characters: a continuous morph from one to the other.

\begin{figure*}[t!]
  \centering
  \includegraphics[height=0.35\textwidth]{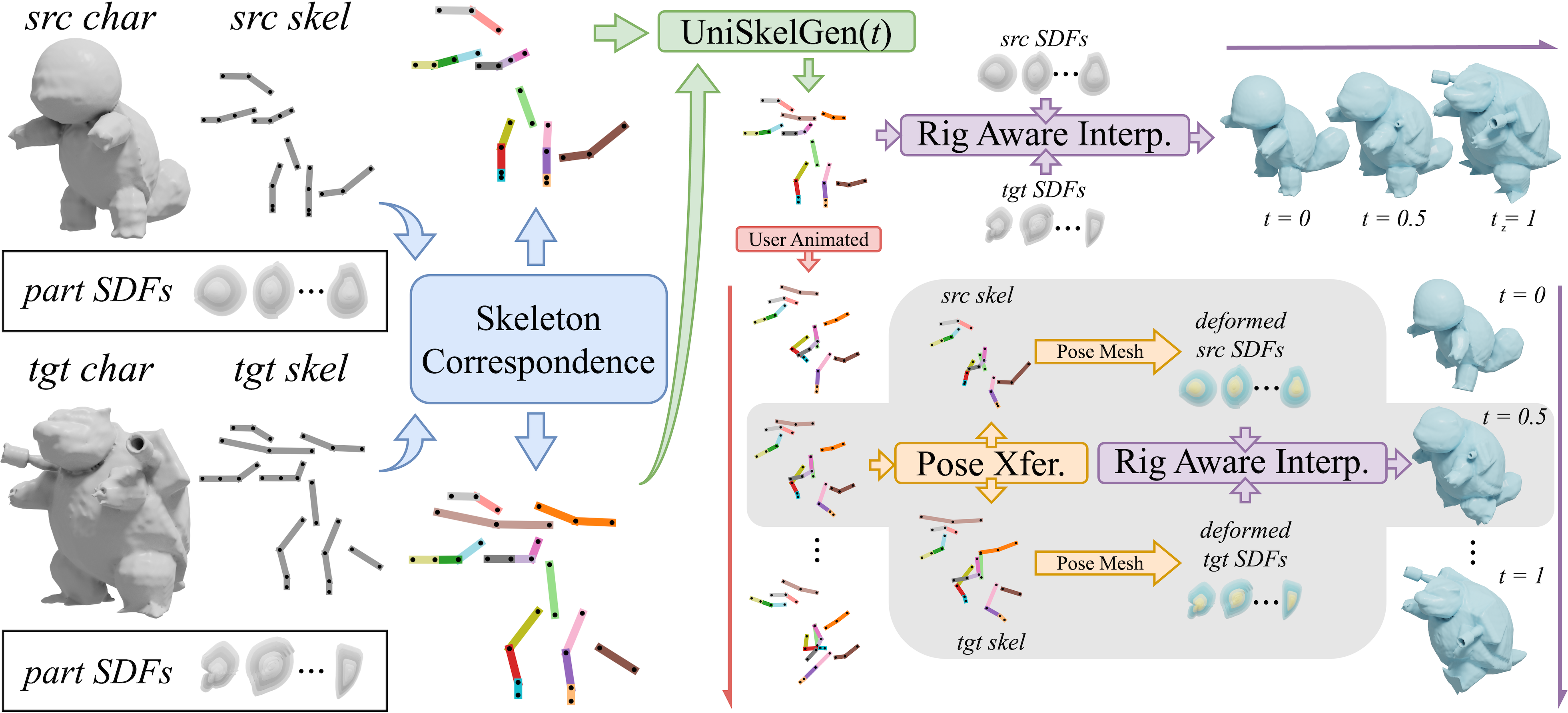}
  \caption{
The overall pipeline of \methodname{}.
Given a source and target character (represented as surface meshes + skeletal rigs), \methodname{} uses rig skinning weights to segment each character's geometry into a set of \textbf{parts}.
It also computes a correspondence between the source and target skeletons, which it uses to create a single \textbf{unified skeleton} given a time step.
This unified skeleton is used to guide interpolation between the geometries of corresponding parts.
Given a posed unified skeleton, \methodname{} transfers the pose to source and target characters and interpolates the deformed geometries of the two posed characters.
\methodname{} enables interpolation during animation, where poses change (red arrow) concurrently with interpolation time steps (purple arrow) as shown in the bottom right.
}
  \label{fig:pipeline}
\end{figure*}

\section{Approach} \label{sec:approach}

Since prior shape interpolation methods~\cite{ConvWasser, NeuroMorph} ignore the underlying structure of shapes, the intermediate results cannot be manipulated.
When the shapes are 3D characters, the interpolated characters are not posable.
\methodname{} resolves this issue by rig-aware interpolation; it maintains a rig throughout the interpolation process. 
Users can pose an intermediate rig to animate an interpolated character.
Furthermore, while the geometry of the unified rig changes with varying interpolation time steps, its topology remains the same. 
Thus, characters can be interpolated for an animation sequence of a single unified skeleton, such that intermediate characters' poses and identities vary at the same time.

Fig.~\ref{fig:pipeline} illustrates \methodname{}'s pipeline. 
Given a pair of source and target characters in rest-pose and their rigs, \methodname{} produces animatable interpolated characters.
\methodname{} first finds hierarchical correspondences between the skeletons of the two characters using recursively defined cost functions (Section~\ref{sec:skelCorr}).
It then creates a \emph{unified skeleton} using the corresponding pairs (Section~\ref{sec:uniSkelGen}).
The unified skeleton serves as a proxy that guides geometry interpolation between topologically-different characters and supports user interaction. 
Users may animate the unified skeleton, and \methodname{} transfers the poses to source and target skeletons by propagating bone transformations (Section~\ref{sec:pose}).
If there is no pose input, the source and target characters remain in their rest poses. 
Lastly, \methodname{} generates geometry for each bone in the unified skeleton by interpolating between the corresponding character part SDFs (Section~\ref{sec:interp}).

\section{Skeleton Correspondence} \label{sec:skelCorr}

Character interpolation should preserve semantics of body parts: legs should be interpolated with legs, and arms should be interpolated with arms, etc.
This calls for a method to find corresponding body parts. 
While existing methods find surface correspondence using functional maps~\cite{FunctionalMaps} or neural networks~\cite{DeepShells, NeuroMorph}, their interpolation ignores rigs such that intermediate results are not posable.
In contrast, we seek to maintain a rig throughout the interpolation process.
Thus, \methodname{} finds hierarchical correspondence between input \emph{skeletons}.
As part of preprocessing, \methodname{} segments rest-pose characters' meshes into surface patches representing body parts using skinning weights, where we assign each vertex to the bone that has the highest skinning weight for that vertex.
If two bones are corresponded, their segmented body parts are also corresponded.
We then convert segmented surface patches of characters in rest-pose to SDFs, which is further discussed in Section~\ref{sec:interp}.

The process of finding bone correspondences is identified as the "Skeleton Correspondence" module in Figure.~\ref{fig:pipeline}.
A bone within a skeleton hierarchy is defined as $(\mathbf{h}_i, l_i, \mathbf{b}_i, M(x_i, y_i, z_i))$, where $\mathbf{h}_i$ is the world position of the head of the bone,  $l_i$ is the bone length, $\mathbf{b}_i$ is the tightest axis-aligned bounding box around the part surface geometry in bone local space, and $M$ transforms from bone local space to world space given by the bone's $x, y, z$ axes. 
Note that the y-axis of bone local space is along the direction of the bone from head to tail, and the x, z axes are computed such that the rotation matrix from the y-axis satisfies the damped track constraint~\cite{DampedTrack}.
$\mathbf{b}_i$'s y-axis is aligned along the bone's y-axis.
The need for $\mathbf{b}_i$ is explained in Section~\ref{subsec:octMap}.
We define source skeleton as bones $\mathbb{S} = \{s_i\}$ and target skeleton as $\mathbb{D} = \{d_j\}$.

\begin{figure}[t!]
  \centering
  \includegraphics[width=0.46\textwidth]{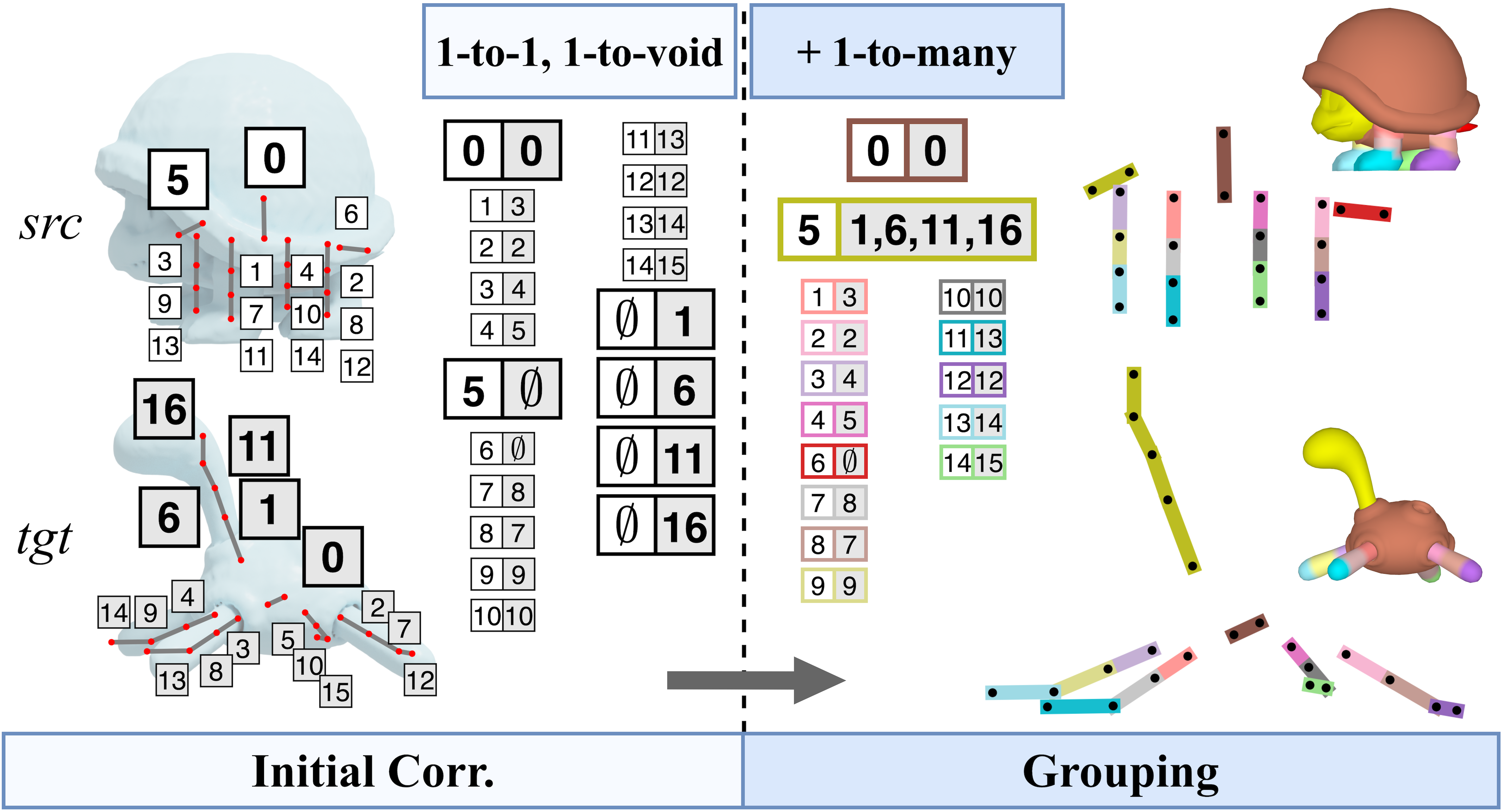}
  \caption{
Our pipeline for computing correspondences between skeletons and constructing an unified skeleton.
Left: All the bones from both source and target skeletons.
Middle: The initial correspondence phase which takes the source skeleton (white) and the target skeleton (gray) and produces 1-to-1 and 1-to-void matches, where void is denoted by $\emptyset$.
Right: Grouping 1-to-void pairs to to create 1-to-many correspondences.
  }
  \label{fig:skelPipeline}
\end{figure}

\methodname{} first produces initial bone correspondences, using Xu et al.~\shortcite{LayoutBlending}'s hierarchical correspondence algorithm with our custom heuristics suitable for 3D bone matching.
The algorithm outputs 1-to-1 and 1-to-void pairs, where one bone can match with another bone or none. 
Having only these two types of pairs is undesirable for 3D skeleton correspondence, as some matching body parts may have different numbers of bones.
\methodname{} addresses this issue by grouping as many 1-to-void correspondences as possible into \emph{1-to-many correspondences}, where one bone is matched to multiple bones. 
In this way, semantically matching body parts with different numbers of bones can be corresponded correctly.
To illustrate, in Fig.~\ref{fig:skelPipeline}, the heads of source and target shapes are correctly corresponded after grouping.

\subsection{Producing Initial Skeleton Correspondences} \label{subsec:initSkelCorr}

\methodname{} finds bone correspondences between two input skeletons to establish part correspondences, and it produces an intermediate character by interpolating geometries of two corresponding parts.
We adapt Xu et al.~\shortcite{LayoutBlending}'s hierarchical correspondence algorithm to find 1-to-1 and 1-to-void bone mappings between two input hierarchies.
Note that a bone in the skeleton hierarchy can be either a leaf bone or a branch bone.
There are five correspondence cases: leaf-to-leaf, leaf-to-void, branch-to-void, branch-to-leaf, and branch-to-branch. 
Xu et al.'s algorithm defines cost functions for leaf-to-leaf and leaf-to-void correspondences, and the costs for the latter three cases are recursively computed from the first two. 
A matrix encoding the cost of matching any source bone to any target bone is then constructed, and the Hungarian algorithm is used to solve for optimal 1-to-1 and 1-to-void correspondences~\cite{Hungarian}. 
The cost function heuristics used in Xu et al.'s algorithm are designed for 2D layouts, so we have developed custom heuristics suitable for 3D skeletons.
In the supplemental material, we enumerate our heuristics and perform an ablation study that validates them.

\subsection{Post-Processing Initial Skeleton Correspondences} \label{subsec:postSkelCorr}

Source and target characters may have semantically corresponding body parts with different numbers of bones.
Fig.~\ref{fig:skelPipeline} shows a source character having a head with one bone (5) and a target character having a head with four bones (1, 6, 11, 16).
When interpolating, since the head bones have 1-to-void correspondence, the source head would shrink and the target head would grow instead of interpolation. 
This is undesirable, so we post-process the initial pairings by introducing 1-to-many correspondences.
A source set and a target set of 1-to-void correspondences can be grouped into 1-to-many correspondence if the lowest common ancestor of the source nodes is in 1-to-1 correspondence with the lowest common ancestor of the target nodes.
In Fig.~\ref{fig:skelPipeline}, the four target bones (1, 6, 11, 16) are grouped together to correspond to the one source bone (5) because all bones satisfy 1-to-void correspondence and the source ancestor (0) is 1-to-1 corresponded with the target ancestor (0).
In the supplemental material, we provide detailed pseudocode for an algorithm that eliminates as many 1-to-void correspondences as possible in favor of 1-to-1 and 1-to-many mappings. 
We define correspondence as $P = (r_s, r_d)$, where $P$ has five cases: $(s, d), (S, d), (s, D), (s, \emptyset), (\emptyset, d)$, $S \subset \mathbb{S}, D \subset \mathbb{D}, s \in \mathbb{S}, d \in \mathbb{D}$.

\section{Unified Skeleton Generation} \label{sec:uniSkelGen}

\begin{figure}[t!]
  \centering
  \includegraphics[width=0.46\textwidth]{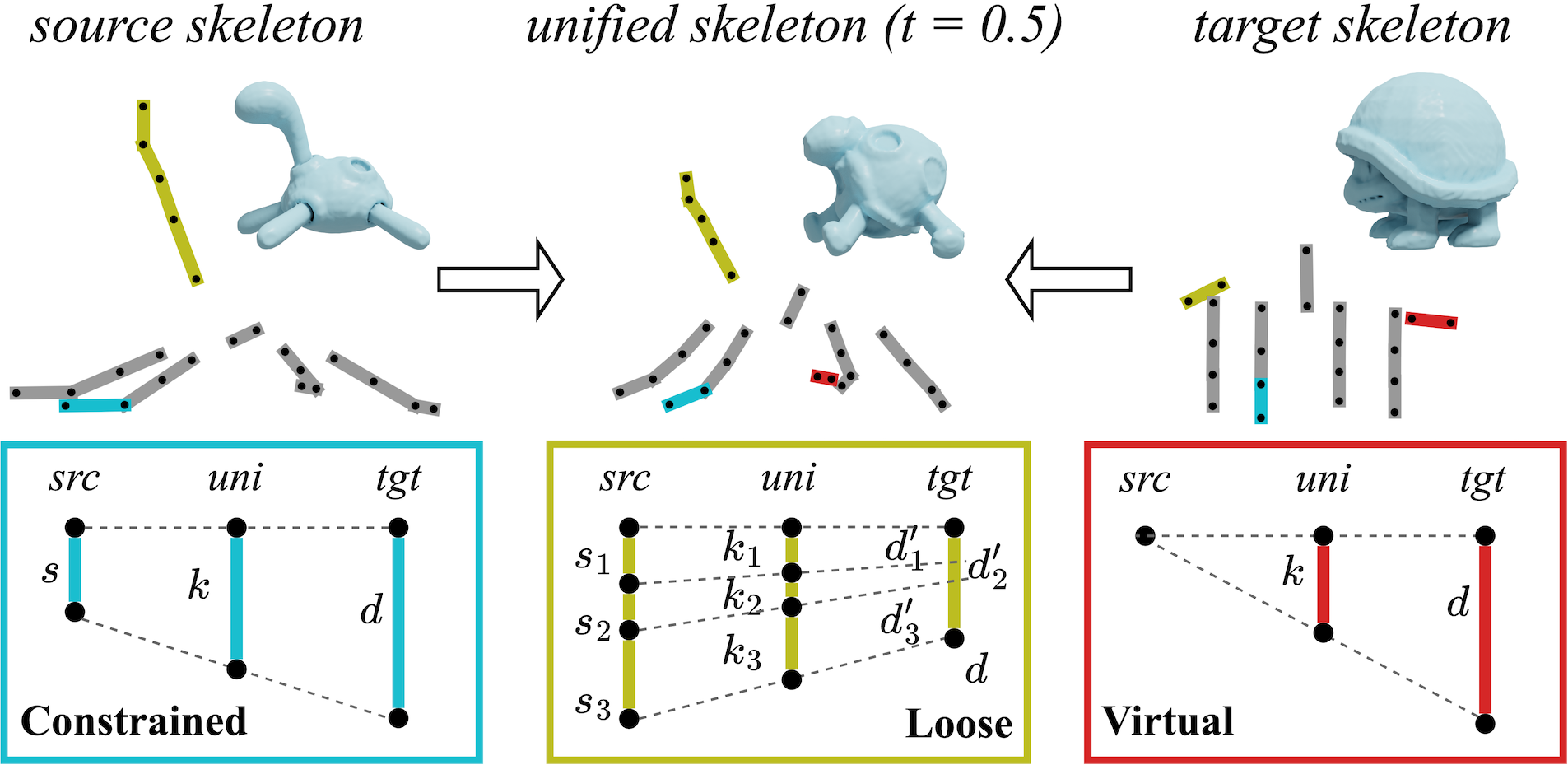}
  \caption{
Generating unified bones at t = 0.5.
The blue claw is a constrained bone made from 1-to-1 correspondence.
Each yellow head bone is a loose bone made from 1-to-many correspondence.
The red tail is a virtual bone made from 1-to-void correspondence.
  }
  \label{fig:uniSkelGen}
\end{figure}

We seek to preserve a rig throughout the interpolation process such that the intermediate characters are posable. 
Given 1-to-1, 1-to-many, 1-to-void skeleton correspondences (Section~\ref{sec:skelCorr}), \methodname{} generates a \emph{unified skeleton} at a time step $t$ (Fig.~\ref{fig:uniSkelGen}).
This process is identified as "UniSkelGen" in Fig.~\ref{fig:pipeline}.
The geometries of unified skeletons vary depending on $t$, but their topologies remain the same.
Thus, animators can interact with a single unified skeleton and specify interpolation steps for each frame to generate an animation sequence where motion and interpolation happen simultaneously (Figs.~\ref{fig:teaser}, ~\ref{fig:interpAnim}).
A unified bone inherits the properties of a regular skeleton bone, and it additionally references source and target bones, a source bone and void, or a target bone and void. 
Define reference $R = (r_s, r_d)$, where $R$ is one of $(s, d), (s, \emptyset), (\emptyset, d), s \in \mathbb{S}, d \in \mathbb{D}$.
A unified bone also carries $t$, denoting the interpolation time step.
Thus, a unified bone is defined as $(\mathbf{h}_k, l_k, \mathbf{b}_k, R_k, t)$, $k \in \mathbb{K}$, where $\mathbb{K}$ is the set of all bones in an unified skeleton.
In this section, we will first discuss the three types of unified bones then explain how to construct a unified bone.
We define three types of unified bones displayed in Fig.~\ref{fig:uniSkelGen}:
\begin{itemize}
\item \textbf{Constrained}. Reference $R = (s, d)$, where $s, d$ satisfy 1-to-1 correspondence $P = (s, d)$.
\item \textbf{Loose}. Reference $R = (s, d)$, where $s, d$ are associated via 1-to-many correspondence $P = (S, d)$ or $(s, D)$, $s \in S, d \in D$.
\item \textbf{Virtual}. Reference $R = (s, \emptyset)$ or $(\emptyset, d)$, where $s, d$ satisfy 1-to-void correspondence $P = (s, \emptyset)$ or $(\emptyset, d)$.
\end{itemize}

Given a 1-to-1 pair $P = (s, d)$, \methodname{} constructs a \textbf{constrained} unified bone (Fig.~\ref{fig:uniSkelGen}, Constrained).
Since we have 1-to-1 mapping, it's straightforward to make an intermediate bone; we linearly interpolate each attribute of $s, d$. 
$\mathbf{h}_k = \text{lerp}(\mathbf{h}_s, \mathbf{h}_d, t)$, $l_k = \text{lerp}(l_s, l_d, t)$, $\mathbf{b}_k = \text{lerp}_\text{box}(\mathbf{b}_s, \mathbf{b}_d, t)$, and $M_k = \text{slerp}(M_s, M_d, t)$, where $\text{lerp}_\text{box}$ linearly interpolates the eight corners of two bounding boxes.
The constrained unified bone references $s$ and $d$. 

Given a 1-to-many correspondence pair $P = (S, d)$, \methodname{} constructs $|S|$ number of \textbf{loose} unified bones (Fig.~\ref{fig:uniSkelGen}, Loose). 
\methodname{} first splits up $d$ into $|S|$ parts whose lengths are proportional to $l_{s_i}, s_i \in S$.
$\mathbf{b}_{k_i}$ is generated similarly by splitting $\mathbf{b}_d$ into parts proportional to the geometries of $\mathbf{b}_{s_i}$ (Section~\ref{subsec:interpWithUniBone}).
We have now converted the problem into 1-to-1 interpolation.
Each $k_i$ can then be constructed in the same way as a constrained bone. 
For example, in Fig.~\ref{fig:uniSkelGen}, $k_2$ is linearly interpolated between $s_2$ and $d'_2$.
The loose unified bone $k_i$ references $s_i$ and $d$, where $s_i \in S$.

Given a 1-to-void correspondence pair $P = (\emptyset, d)$, \methodname{} constructs a \textbf{virtual} unified bone (Fig.~\ref{fig:uniSkelGen}, Virtual).
\methodname{} uses \emph{bounding box mapping} to map $\mathbf{h}_d$ to the source character, where the details of bounding box mapping are explained in Section~\ref{subsec:octMap}.
Note that the center of the bounding box sits at the head of the root bone, and the axes of the box are aligned with those of the root bone. 
Then, \methodname{} linearly interpolates $\mathbf{h}_d$ and the projected point to acquire $\mathbf{h}_k$.
The virtual unified bone $k$'s length and bounding box is computed by $l_k = \text{lerp}(0, l_d, t)$, and $\mathbf{b}_k = \text{lerp}_\text{box}(\mathbf{b}_0, \mathbf{b}_d, t)$, where $\mathbf{b}_0$ denotes a bounding box whose height, width, and length are all 0.
The virtual unified bone references $\emptyset$ and $d$. 
The supplemental material contains more detailed pseudocode for the unified skeleton construction process.

\section{Pose Transferring} \label{sec:pose}

Users can interact with the unified skeletons to pose or animate interpolated characters.
\methodname{} also produces animation sequences where interpolation happens simultaneously as shown in Fig.~\ref{fig:teaser}.
Given animation input for a unified skeleton at $t$, \methodname{} transfers poses to source and target characters' skeletons and blends their geometries to generate a posed or animated interpolated character at interpolation time $t$. 
Note that by construction, even though the geometry of the unified skeleton varies depending on time step $t$, its topology remains the same throughout the interpolation.
Thus, users can animate any unified skeleton and also vary the interpolate time step such that characters are interpolated during an animation sequence. 
In this section, we will discuss how pose is transferred from a unified skeleton to source and destination skeletons.
This process is identified as "Pose Xfer." in Fig.~\ref{fig:pipeline}.

\methodname{} transfers the unified skeleton's pose to source and target skeletons by propagating bone transformations.
We will explain how \methodname{} transfers poses from constrained, loose, and virtual unified bones.

A \textbf{constrained} unified bone $k$ references one source bone $s$ and one target bone $d$, where $s$ and $d$ have 1-to-1 correspondence. 
In this case, the source and target nodes should transform in the same way as the unified bone.
Given user input joint angles for the unified bone, \methodname{} computes a local rotation matrix $Rot_k$ and sets $Rot_s = Rot_d = Rot_k$.

\textbf{Loose} unified bones are created from 1-to-many correspondence $P = (S, d)$, and each unified bone $k_i$ references $R_i = (s_i, d), s_i \in S$.
When users pose $k_i$, the behavior of $s_i$ should be the same; ${Rot}_{s_i} = {Rot}_{k_i}$.
To rotate bone $d$, one way would be to compute the vector from the head of the $S$ linkage to its tail and rotate bone $d$ to align with that vector.
However, this would ignore the rotation around the axis along bone $d$.
Thus, \methodname{} averages the joint angles of $\{k_i\}$ to construct rotation matrix $Rot_d$.

A \textbf{virtual} unified bone $k$ references $R = (s, \emptyset)$ or $R = (\emptyset, d)$.
If $R = (s, \emptyset)$, rotating $k$ only affects $s$ and has no impact on the target skeleton. 
Thus, \methodname{} sets $Rot_s = Rot_k$ or $Rot_d = Rot_k$.

\section{Rig-Aware Geometry Interpolation} \label{sec:interp}

We have discussed how \methodname{} computes skeleton correspondence and constructs a unified skeleton. 
It is established that bone correspondence implies part correspondence (Section~\ref{sec:skelCorr}), and each unified bone refers to a source and/or a target bone in either 1-to-1, 1-to-many, or 1-to-void correspondence (Section~\ref{sec:uniSkelGen}).
The unified skeleton allows for rig-aware character interpolation; a rig is maintained throughout interpolation, and \methodname{} builds geometries around the unified skeleton.
Given user pose or animation input for a unified skeleton, \methodname{} transfers the pose to source and target characters' skeletons and interpolates their posed geometries. 
If there is no pose input, \methodname{} interpolates their rest-pose geometries.
Note that although the geometries of the unified skeletons vary depending on the input interpolation time step $t$, their topologies remain the same. 
Thus, users can pose a single unified skeleton generated at an arbitrary $t$ and pass different interpolation steps to \methodname{}'s rig-aware geometry interpolation algorithm for each frame of animation.
\methodname{} then produces an animation sequence where interpolation and motion happen at the same time (Fig.~\ref{fig:teaser} bottom row).
In Fig.~\ref{fig:pipeline}, a user animates the unified skeleton generated at $t=0.5$ (center) and passes different interpolation steps for each animated frame to \methodname{} and produces a run cycle sequence where the source character is interpolated to the target character.

\begin{figure}[t!]
  \includegraphics[width=0.46\textwidth]{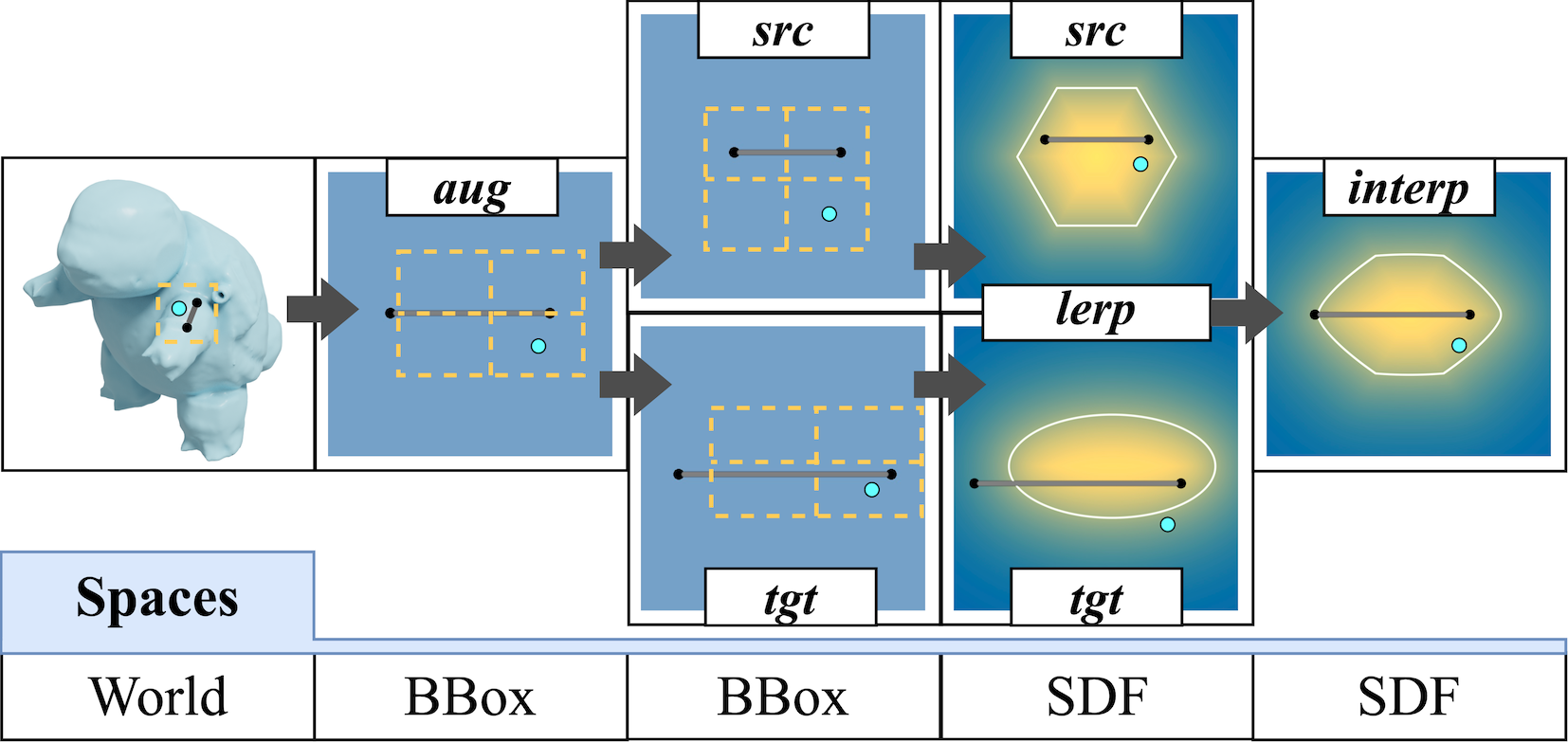}
  \caption{
Interpolating part geometries for a constrained unified bone.
With part-based SDFs, interpolating between two characters amounts to taking a query point (cyan dot) within the bounding box of a unified skeleton bone in the bone's local space, finding which point in the source and target boxes that it maps to, looking up SDF values at those points, and linearly interpolating between them.
  }
  \label{fig:geomInterp}
\end{figure}

Since interpolation of source and target geometries should preserve part-level semantics (e.g. legs should be interpolated with legs), \methodname{} employs part-based representation of characters, and \methodname{} uses each unified bone's reference information $R$ to identify and interpolate source and target parts.
We choose signed distance fields (SDFs) to represent characters' body parts, as SDFs allow easy interpolation between geometries with different mesh topologies.
Furthermore, \methodname{}'s geometry interpolation is aware of the rig state, which allows for interpolating character geometries during animation,; \methodname{} keeps track of bone local frames and interpolates parts in bone local spaces where SDFs are defined.
To obtain an SDF for a segmented body part, \methodname{} converts the part to voxels on a $128^3$ regular grid then to SDFs using distance transform.
When interpolating part SDFs, \methodname{} uses \emph{bounding box mapping} to ensure that the interpolated geometry preserves characteristics from both parts.
Fig.~\ref{fig:geomInterp} shows how \methodname{} acquires SDF values for a blended part by mapping a query point to the source and target spaces then interpolating.
A query point is evaluated by each unified bone $k \in \mathbb{K}$ to produce SDF value $v_k$; \methodname{} defines the final SDF value for that point as $V_k = \min\{v_k\}$, which can be interpreted as a union operation.
In the following subsections, we explain bounding box mapping, discuss shape interpolation for different types of unified bones, and show how to query deformed SDFs given character poses.

\subsection{Bounding Box Mapping} \label{subsec:octMap}

\begin{figure}[t!]
  \centering
  \includegraphics[width=0.46\textwidth]{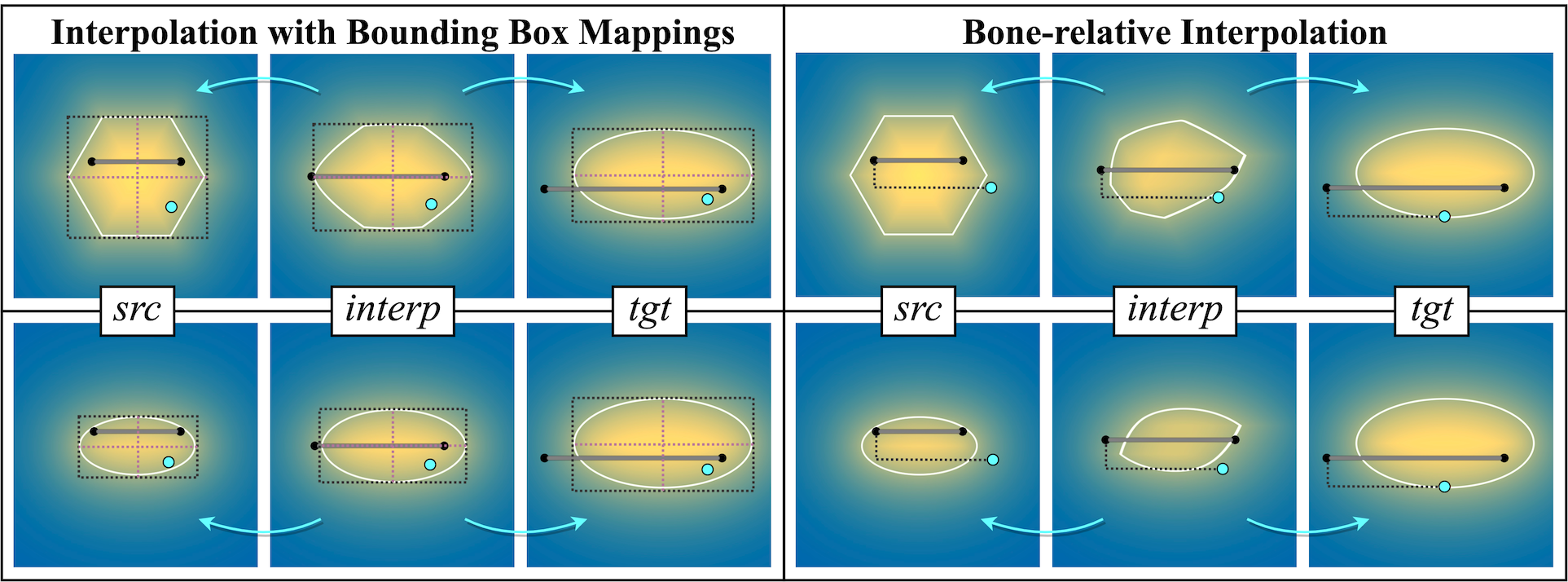}
  \caption{
Comparison between interpolation using bounding box mapping and bone-relative mapping with 2D SDF visualization, where white outlines geometries and grey denotes bones.
The interpolated geometry (``interp") is generated by mapping a point (cyan dot) in the unified bone's space to source and target spaces and then interpolating the queried SDFs. 
Interpolation with bounding box mapping preserves characteristics of input geometries while the other method does not.
  }
  \label{fig:octantMapping}
\end{figure}

When interpolating between corresponding parts, the intermediate result should preserve characteristics from both parts.
\methodname{} achieves this by re-localizing query points with bounding box centers and scaling them when transforming to source and target spaces.
Fig.~\ref{fig:octantMapping} shows two methods for part interpolation.
If we map a query point $p$ in the unified bone space to source and target spaces relative to its position to the bone heads, interpolating a hexagon and an ellipse results in an unexpected shape.
In contrast, \textit{bounding box mapping} outputs a rounded hexagon. 
\methodname{} first constructs bounding boxes $B_s, B_d$ around source and target part mesh geometries and interpolate to obtain an intermediate bounding box $B_k = \text{lerp}_\text{box}(\mathbf{b}_s, \mathbf{b}_d, t)$ for a constrained or loose unified bone $k$.
Then \methodname{} finds $p$'s location in the intermediate bounding box before scaling it to source and target bounding box spaces.
Let $\mathbf{b}^c$ denotes the center of $\mathbf{b}$, and $\mathbf{b}^x, \mathbf{b}^y, \mathbf{b}^z$ denotes the x, y, z axes lengths of $\mathbf{b}$.
Point $p$ in the source bounding box space is given by $(p - \mathbf{b}^c_k) \cdot (\frac{\mathbf{b}^x_s}{\mathbf{b}^x_k}, \frac{\mathbf{b}^y_s}{\mathbf{b}^y_k}, \frac{\mathbf{b}^z_s}{\mathbf{b}^z_k})$.
Note that a query point may land outside of the interpolated bounding box, but the transformation still applies as bounding box geometries provide scaling.

\subsection{Interpolation with Unified Bones} \label{subsec:interpWithUniBone}

\methodname{} uses a unified bone $k$'s reference information $R$ to interpolate between source and destination part geometries.

It's simple to compute geometry for a \textbf{constrained} unified bone $k$ which references one source bone $s$ and one target bone $d$. 
Fig.~\ref{fig:geomInterp} shows how to obtain an interpolated SDF value given a query point $p$ in world space. 
\methodname{} first transforms $p$ into the interpolated bounding box $\mathbf{b}_k = \text{lerp}_\text{box}(\mathbf{b}_s, \mathbf{b}_d, t)$.
Using bounding box mapping, \methodname{} finds the position of $p$ in $\mathbf{b}_s$, $\mathbf{b}_d$. 
SDF values of source and target parts are defined in bone local space, so \methodname{} transforms $p$ from bounding box space to bone local space to query $v_s, v_d$, where the point is translated by $\mathbf{b}^c$, the bounding box center.
The final SDF value is given by $v_k = \text{lerp}(v_s, v_d, t)$.

When source and target bones are in 1-to-many correspondence $P = (S, d)$, \methodname{} constructs $|S|$ number of \textbf{loose} unified bones $k_i$, each referencing $R_i = (s_i, d), s_i \in S$, where each $s_i$ has bounding box ${\mathbf{b}_s}_i$.
Similar to how \methodname{} proportionally splits up bone $d$ when generating $k_i$, \methodname{} splits $d$'s bounding box $\mathbf{b}_d$ into sub-boxes $\mathbf{b}'_{d_i}$ whose y-axis lengths are proportional to each of $\mathbf{b}_{s_i}$.
In this way, an interpolated bounding box can be generated for each loose unified bone $k_i$, where $\mathbf{b}_{k_i} = \text{lerp}_\text{box}(\mathbf{b}_{s_i}, \mathbf{b}'_{d_i}, t)$.
\methodname{} then proceeds to interpolate geometries as described in the previous paragraph.

A \textbf{virtual} unified bone $k$ references $(s, \emptyset)$ or $(\emptyset, d)$.
  Interpolation for virtual bones works similarly to constrained bones.
When either the source or destination bounding box is $\mathbf{b}_0$, \methodname{} doesn't attempt to transform point $p$ into $\mathbf{b}_0$ but sets $v_s = 0$ or $v_d = 0$ then interpolates for $v_k$.

\subsection{Querying Deformed SDFs} \label{subsec:deformedSDF}

Given a posed unified skeleton and an interpolation time step, \methodname{} queries deformed SDFs of source and target characters to generate an interpolated character.
After transferring the unified skeleton's pose to source and target skeletons, \methodname{} poses the two input characters using their respective skinning weights.
\methodname{} then segments the characters into parts using the procedure described in the beginning of Section~\ref{sec:interp}, converts them to SDFs, and uses rig-aware interpolation (Section~\ref{subsec:interpWithUniBone}) to generate geometry for the posed unified skeleton.

\begin{figure*}[t!]
  \centering
  \includegraphics[width=0.83\textwidth]{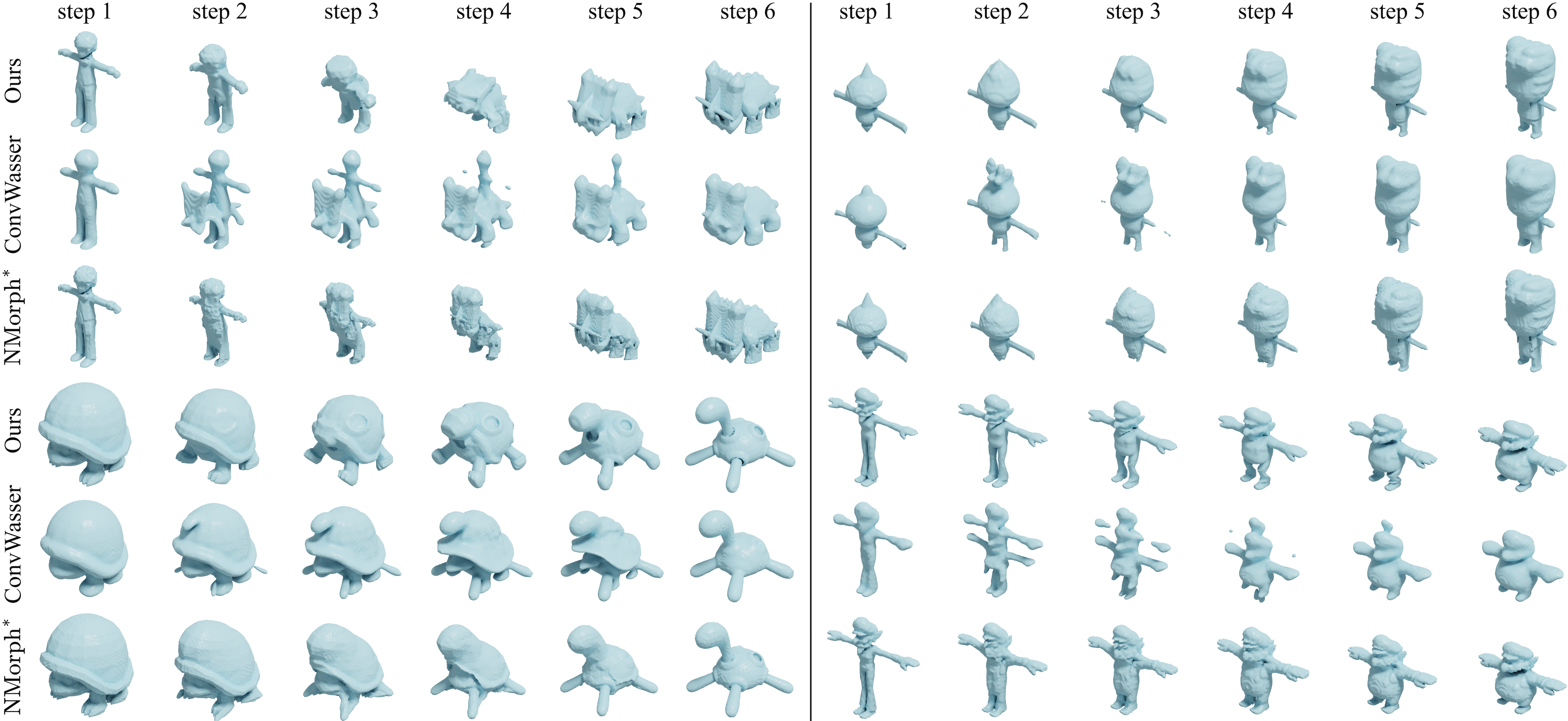}
  \caption{
    Qualitative comparisons between \methodname{}, ConvWasser~\cite{ConvWasser}, and NeuroMorph~\cite{NeuroMorph} (``$\text{NMorph}^*$").
    The ``$\text{NMorph}^*$" rows are produced using the method in Section~\ref{sec:resultsAndEval}.
    \methodname{} smoothly interpolates characters while the others do not.
    ConvWasser is agnostic to correspondences, and NeuroMorph produces many artifacts (arms at steps 3, 4 of top left; head and paws at steps 3, 4 of bottom left; legs at steps 2, 3, 4 of top right; arms and head at steps 2, 3, 4, 5 of bottom right). 
    In the supplemental document, we show two sets of NeuroMorph interpolations generated from both directions for each pair. See supplementary video for animated comparisons. }
  \label{fig:moreInterpComp}
\end{figure*}

\section{Results and Evaluation} \label{sec:resultsAndEval}

In this section, we present results of interpolating a variety of characters with \methodname{}.
We compare \methodname{} with ConvWasser~\cite{ConvWasser}, a rig-oblivious method for shape interpolation with optimal transport, and NeuroMorph~\cite{NeuroMorph}, a learning-based approach that produces surface correspondence and interpolation given two input meshes.
We trained each character pair with NeuroMorph for 5000 epochs to obtain results.
For our experiments, we used 36 character pairs from the RigNet dataset~\cite{RigNet}, sourced from Models Resource~\cite{ModelsResource}, a publicly-available dataset of rigged video game characters.
Experiments were run on 6-core Intel i7 machine with 32GB RAM and a NVIDIA GTX 1080Ti GPU.
We strongly encourage readers to watch our supplementary video for animated results.

\begin{figure}[t!]
  \centering
  \includegraphics[width=0.47\textwidth]{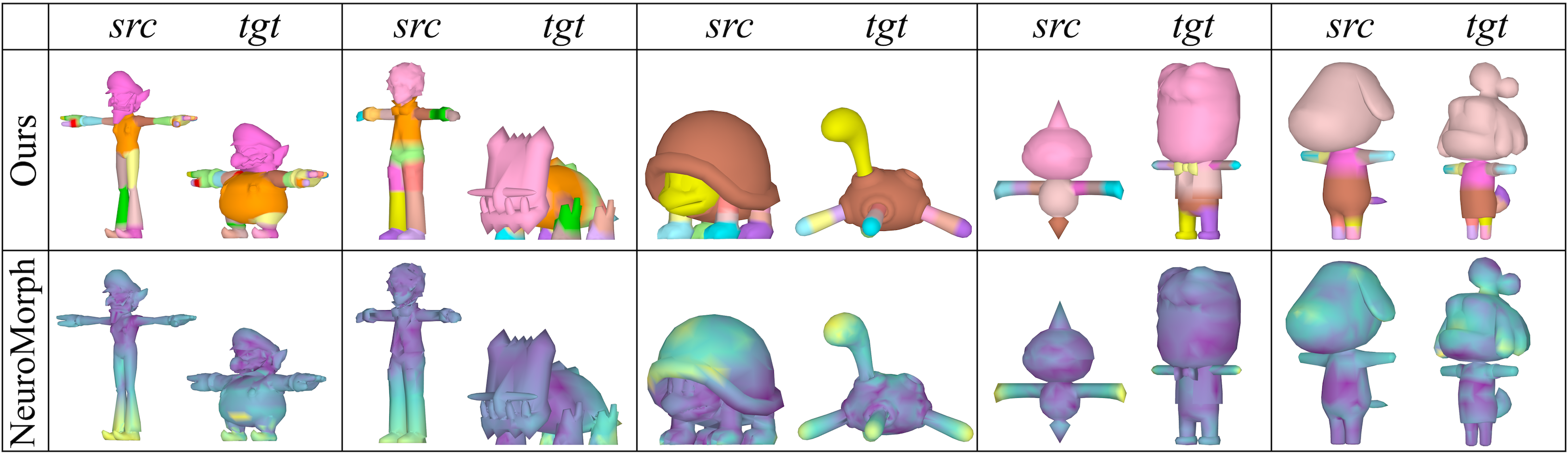}
  \caption{
Qualitatively comparing our correspondences with NeuroMorph.
We visualize \methodname{}'s skeleton correspondence by segmenting meshes using skinning weights and assigning different colors to surface patches associated with corresponding bones.
NeuroMorph colors denote vertex-level correspondence.
}
  \label{fig:skelCorresResults}
\end{figure}

Fig.~\ref{fig:skelCorresResults} compares \methodname{} and NeuroMorph correspondences.
Leveraging characters' rigs, \methodname{} produces more fine-grained correspondences and correctly identifies corresponding body parts.
For instance, in Fig.~\ref{fig:skelCorresResults} column 3, we match the two heads of the character pair, while NeuroMorph matches the shell of the first character to the head of the second character.
See the supplemental material for more of our correspondences.

Fig.~\ref{fig:moreInterpComp} compares how well \methodname{} interpolates rest-pose characters compared to ConvWasser and $\text{NeuroMorph}^*$. 
NeuroMorph changes the pose of one shape to match the other while leaving its identity unchanged.
We produced NeuroMorph's interpolation for both directions from source $s$ to target $d$ and $d$ to $s$ to obtain ${s_{t=0}, ... s_{t=1}}$ and ${d_{t=0}, ... d_{t=1}}$. 
Intermediate results, labeled $\text{NeuroMorph}^*$, are computed by $\text{lerp}(\text{SDF}(s_{t}), \text{SDF}(d_{1-t}), t)$ where $\text{SDF}(\cdot)$ converts a mesh to SDF.
Our approach produces higher-quality and semantic-preserving interpolations. 
For example, in the bottom right of Fig.~\ref{fig:moreInterpComp}, our approach smoothly interpolates the heads, while ConvWasser ignores correspondence and $\text{NeuroMorph}^*$ has many artifacts around the intermediate characters' head and arms.
Furthermore, our method allows intermediate characters to be posed, while ConvWasser and NeuroMorph do not.

\begin{figure*}
  \centering
  \includegraphics[width=0.85\textwidth]{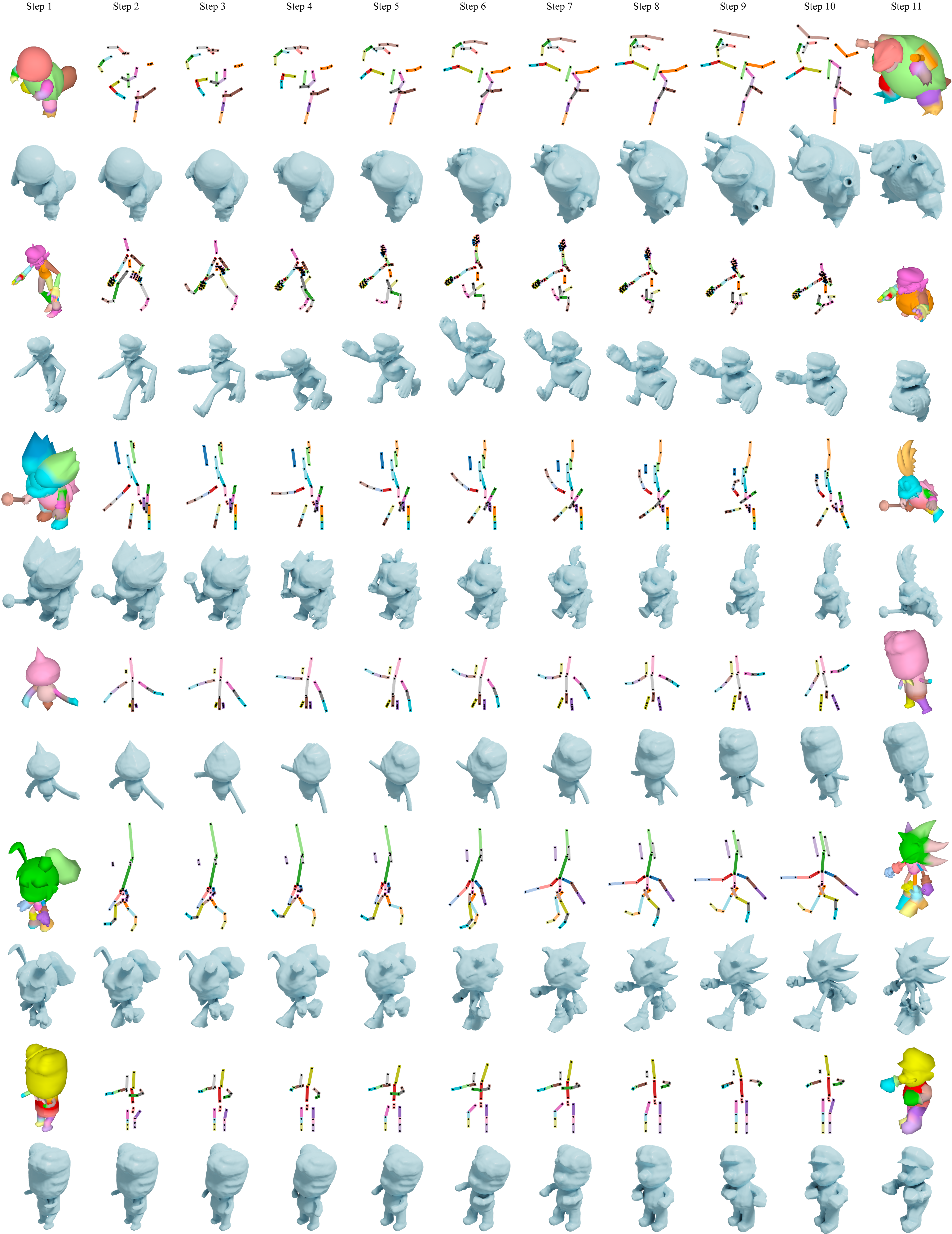}
  \caption{
  Interpolation during animation.
  For each character pair, the top row shows correspondence and posed unified skeletons, and the row below shows a motion sequence where character poses and interpolation steps vary at the same time.
  During interpolation, \methodname{} preserves a ``unified skeleton" whose geometry varies depending on time steps and topology remains the same.
  Thus, animators can interact with a single unified skeleton and specify interpolation steps for each frame to achieve interpolation during animation.
  Pairs 1, 4 include 1-to-1, 1-to-void, and 1-to-many correspondences, and pairs 2, 3, 5, 6 include 1-to-1 and 1-to-void correspondences.
  Please watch our supplementary video for animated sequences. 
  }
  \label{fig:interpAnim}
\end{figure*}

Fig.~\ref{fig:interpAnim} shows interpolation during animation. 
For each sequence, an artist has animated the unified skeleton generated at $t = 0.5$ (Section~\ref{sec:uniSkelGen}) and specified interpolation time steps for each frame. 
\methodname{} then transfers the poses to source and target characters (Section~\ref{sec:pose}) and performs rig-aware interpolation (Section~\ref{sec:interp}).
Since \methodname{} maintains a rig with the same topology throughout interpolation, it allows for interpolation while the intermediate characters perform a motion sequence.

In terms of timing, \methodname{} produces a posed interpolated character in 83 seconds on average. 
We experimented with another approach to generate geometry for a posed intermediate character that is 17\% faster with negligible cost in quality (see the supplemental material). 

\section{Conclusion}

We presented \methodname{}, a method to interpolate between 3D characters with different mesh and rig topologies, such that users can pose the intermediate interpolated characters. 
It also enables interpolation during animation.
To the best of our knowledge, \methodname{} is the first system that tackles this novel challenge.
\methodname{} achieves this goal by maintaining a unified rig throughout interpolation, where the unified rig is built from skeleton correspondences between two input rigs of potentially different topologies.
\methodname{} is agnostic to mesh topologies as it represents characters as a union of sign distance fields, one per each bone of the character's rig.
We showed how to perform rig-aware interpolation of characters and pose any intermediate interpolated character.
Our experiments show that \methodname{} produces higher-quality character interpolations than rig-oblivious shape interpolation methods~\cite{ConvWasser, NeuroMorph}.

\methodname{} is not without its limitations.
Similar to Xu et al.~\cite{LayoutBlending}'s work, our skeleton correspondence algorithm can sometimes produce incorrect correspondences that may not satisfy users.
In this case, an interactive system built on \methodname{} could simply allow users to manually correct the automatically-produced correspondences.
Moreover, \methodname{} can struggle to produce a good correspondence between characters with drastically different skeletons---indeed, in some cases, a meaningful correspondence might not exist.
In the supplemental material, we provide some correspondence failure cases.
Although we have shown that NeuroMorph~\cite{NeuroMorph}, a mesh-based method,  creates severe artifacts and does not support interpolation during animation, \methodname{}'s SDF-based representation has a lack of control over output surfaces such as loss of original mesh topology and texture.
In terms of posing intermediate results by propagating poses to source and target characters, the current pose transferring method works well when the local frames of matching bones align well.
It is an interesting future direction to implement more dedicated pose retargeting to refine the intermediate characters' motion.
Nevertheless, such dedicated techniques only need to deal with local motion retargeting as we have simplified the problem to three categories – constrained, loose, and virtual.
Additionally, more work is needed to optimize \methodname{} for real-time use. 
To pose an interpolated character, 70\% of reconstruction time is spent on converting part-level mesh geometries to SDFs with voxelization and distance transform.
The runtime can be improved by employing faster methods to convert from mesh to SDFs, such as fast winding numbers~\cite{FastWindingNumbers}.
Neural SDFs~\cite{NeuralCompact} may further increase the speed of interpolation.

\section{Acknowledgements}

We would like to thank Ivery Chen and Healey Koch for animating interpolation sequences, and Arman Maesumi for providing radial basis function interpolator code for our fast interpolation by advection  method (Supplemental Section 3).

\bibliographystyle{eg-alpha} 
\bibliography{main}       

\end{document}


\maketitle

\section{Skeleton Correspondence}

We adapted ~\cite{LayoutBlending}'s hierarchical correspondence algorithm to compute 1-to-1 and 1-to-void correspondence.
The cost functions in Xu et al.'s work are designed for 2D layouts, so we have implemented our custom cost functions for 3D skeleton matching.
Furthermore, we post-process the 1-to-1 and 1-to-void correspondences by grouping as many 1-to-void pairs to 1-to-many pairs as possible, as semantically corresponding body parts may have different numbers of bones.
Note that correspondence is computed with character skeletons in rest-pose as the artist-authored rigs are in rest-pose by default.
In this section, we enumerate our heuristics (Section~\ref{subsec:heuristics}), present pseudocode for our post-processing method (Section~\ref{subsec:postProcess}), and and perform an ablation study (Section~\ref{subsec:ablation}).

Note that a related work SkinMixer~\cite{SkinMixer} also employs heuristics to find bone-to-bone correspondence before mixing and matching parts from rigged character models.
However, \methodname{} uses different heuristics and utilizes recursion with the Hungarian algorithm~\cite{Hungarian} rather than propagation to compute correspondences.

\subsection{Cost Function Heuristics} \label{subsec:heuristics}
A node in a hierarchy can be either a leaf node or a branch node.
There are five correspondence cases: leaf-to-leaf, leaf-to-void, branch-to-void, branch-to-leaf, and branch-to-branch. 
Xu et al.'s algorithm defines cost functions for leaf-to-leaf and leaf-to-void correspondences, and the costs for the latter three cases are recursively computed from the first two. 
In addition, we have found that adding an additional term to the branch-to-branch cost improves correspondence accuracy. 
We list our heuristics below.

\subsection{Leaf-to-leaf}
\begin{figure}[t!]
  \centering
  \includegraphics[width=0.47\textwidth]{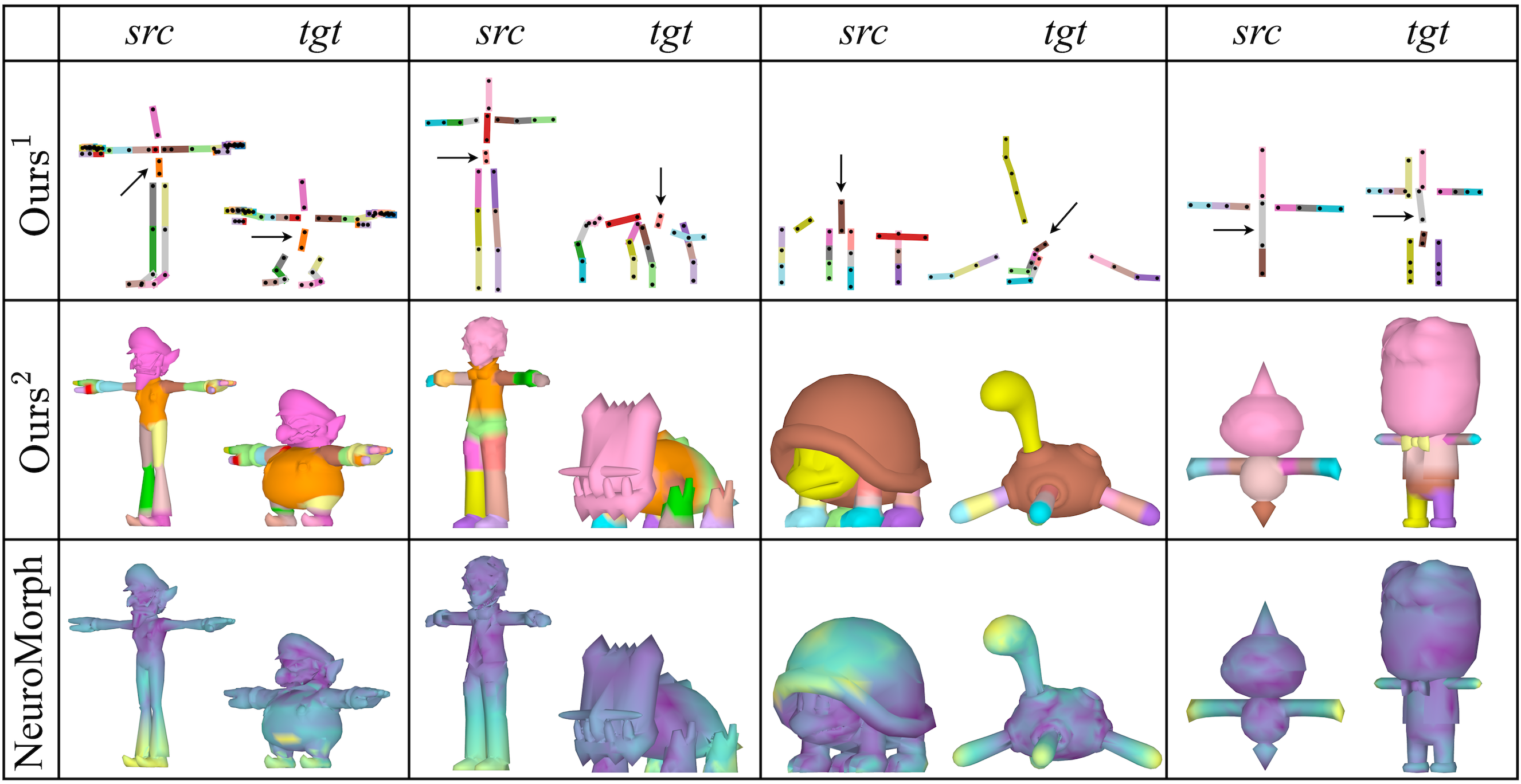}
  \caption{
Qualitatively comparing our correspondences with NeuroMorph~\cite{NeuroMorph}.
``$\text{Ours}^1$" shows corresponding skeletons where root bones are indicated by arrows.
``$\text{Ours}^2$" visualizes skeleton correpondence by segmenting each mesh using skinning weights and assigning different colors to surface patches associated with corresponding bones.
  }
  \label{fig:CorrComp_supp}
\end{figure}

For a source leaf bone $s$ and a target leaf bone $d$, we formulate the leaf-to-leaf cost as
\begin{equation}
\label{eqn:ltl}
    C_{l-l}(s, d) =
    |l_s - l_d| +
    || \mathbf{h}_s - \mathbf{h}_d ||_2 +
    |j_s - j_d| +
    \mathds{1}_{O(\mathbf{h}_s) = O(\mathbf{h}_d)}
\end{equation}
where $j$ is the hierarchy level of the bone, and $O(q)$ returns the octant where $q$ lies. 
Note that the octants are centered around the head of the root bone $r$, and their axes are aligned with $x_r, y_r, z_r$. 
The first two terms in the cost function indicate difference in bone geometry; bones that are similar in length and world position should have low cost.
The last two terms indicate difference in bone semantics.
First, bones with different levels in the hierarchy should have a higher cost.
Second, 3D characters often have skeletons of very different geometries.
Consider matching the arms of a bipedal to the front legs of a quadruped in Fig.~\ref{fig:CorrComp_supp}, column 2.
The arm and front leg bones are in very different locations in world space.
However, the arm and front leg bones are in the same octant relative to the root bone.
See Section~\ref{subsec:ablation} for an ablation study that validates the use of these four heuristics.

\subsection{Leaf-to-void} Similar to Xu et al., we define this cost as the effort for deforming the leaf bone to void.
We formulate the cost as
$
    C_{l-\emptyset}(n) = \alpha * L_n
$.
We have observed that $\alpha$ has a significant impact on correspondence accuracy, so we developed a heuristic to compute a suitable $\alpha$ for a given pair of skeletons:
\begin{displaymath}
    \alpha(S, D) = c_1  \text{min}(|S|, |D|) + c_2  ||S| - |D|| + c_3
\end{displaymath}
where parameters $c_1 = -0.05, c_2 = 0.15, c_3=1.5$ are determined by running multi-linear regression on 15 pairs of characters with manually labeled correspondences.
We consider the minimum number of source and target bones because we observe that when there are more bones to match, more bones should be allowed to match with void. 
Leaf-to-void cost matters the most when two inputs skeletons don't have the same number of bones, so we also consider the difference in the number of source and target bones in our heuristics.
This predictor achieves 86\% correspondence accuracy on a held-out set of 36 diverse character pairs.

\subsection{Branch-to-branch} 
Semantically similar branch bones often have similar directions to their roots.
For instance, the direction from the wrist bone to the root is different than that from the ankle bone, so the wrist bones from two characters should match (Fig.~\ref{fig:CorrComp_supp}, column 1).
After recursively computing the cost between two branch bones $s$ and $d$, we add an additional term formulated as
\begin{equation}
\label{eqn:btb}
    C_{b-b}(s, d) = \frac{w_s \cdot w_d}{\lVert w_s \rVert \lVert w_d \rVert}
\end{equation}
where $w$ is the direction from the head of a bone to the head of the skeleton's root bone.
The ablation study in Section~\ref{subsec:ablation} shows that this cost is essential to increase correspondence accuracy.

\subsection{Post Processing Initial Correspondence} \label{subsec:postProcess}

\begin{algorithm}[t!]
\SetAlgoLined
\LinesNumbered
\caption{Post-Processing Initial Skeleton Correspondences}
\label{algo:postProcess}
\KwIn{1-to-1, 1-to-void correspondence pairs}
\KwOut{1-to-1, 1-to-void, 1-to-many correspondences pairs}
\textit{\textbf{o\_pairs}} $\leftarrow$ 1-to-1 pairs; \textit{\textbf{v\_pairs}} $\leftarrow$ 1-to-void pairs \\
\textit{src\_void\_bones}, \textit{tgt\_void\_bones} $\leftarrow$ all src, tgt bones that map to void \\
\textit{\textbf{m\_pairs}} = [] \tcp{1-to-many pairs}
\For{(\text{src}, \text{tgt}) \text{in} \textit{o\_pairs}}{
    \textit{src\_void\_chains}, \textit{tgt\_void\_chains} $\leftarrow$ all chains $B$ of \textit{src}, \textit{tgt} s.t. all bones on $B$ map to void \tcp*{$B$: a chain of bones where each bone can have at most one child}
    \textit{src\_hs}, \textit{tgt\_hs} $\leftarrow$ ancestors of \textit{src\_void\_brchs}, \textit{tgt\_void\_brchs} \tcp{ancestor: bone in a chain with the highest hierarchy level} 
    $C$ $\leftarrow$ cost matrix constructed by computing pair-wise branch-to-branch cost between \textit{src\_hs}, \textit{tgt\_hs} \\
    \textit{chain\_pairs} $\leftarrow$ linear\_sum\_assignment($C$) \\
    \For{(\textit{src\_chain}, \textit{tgt\_chain}) \text{in} \textit{chain\_pairs}}{
        \textit{s, d} $\leftarrow$ all source and target bones in \textit{src\_chain}, \textit{tgt\_chain} \\
        \textbf{remove} pairs from \textit{\textbf{v\_pairs}} that involve \textit{s, d} \\
        \eIf{either \textit{src\_chain} or \textit{tgt\_chain} only has one bone}{
            \textbf{add} (\textit{src\_chain}, \textit{tgt\_chain}) to \textit{\textbf{m\_pairs}}
        }{
            \tcp{src\_chain, tgt\_chain both have >=1 bones} 
            n $\leftarrow$ min(len(\textit{src\_chain}), len(\textit{tgt\_chain})) - 1 \\
            \For{i in range(n)}{
                \textbf{add} (\textit{src\_chain}[i], \textit{tgt\_chain}[i]) to \textit{\textbf{o\_pairs}} \\
                remove \textit{src\_chain}[i] from \textit{src\_chain}, \textit{tgt\_chain}[i] from \textit{tgt\_chain}
            }
        \textbf{add} the remaining (\textit{src\_chain}, \textit{tgt\_chain}) to \textit{\textbf{m\_pairs}}
        }
    }
}
\end{algorithm}

Using our custom cost functions with ~\cite{LayoutBlending}'s hierarchical correspondence algorithm, we obtain 1-to-1 and 1-to-void correspondences.
Since semantically corresponding body parts may have different numbers of bones (Fig.~\ref{fig:CorrComp_supp}, column 3, heads), we post-process the initial correspondences by grouping 1-to-void correspondences to as many 1-to-many pairs as possible. 
Algorithm~\ref{algo:postProcess} shows pseudocode for our algorithm that post-processes initial skeleton correspondences to introduce 1-to-many correspondences (introduced in Section 4.2 of the main paper).
The intuition is that a source set and a target set of 1-to-void correspondences can be grouped into 1-to-many correspondence if the lowest common ancestor of the source bones is in 1-to-1 correspondence with the lowest common ancestor of the target bones.

Given a pair of bones in 1-to-1 correspondence, we find the chains under \textit{src}, \textit{dest} where each bone on the chain has 1-to-void correspondence.
Then, we use the \textit{branch-to-branch} cost listed in Section~\ref{subsec:heuristics} to construct a cost matrix to match the chains.
For a pair of chains, we identify the source and target bones and remove all 1-to-void correspondence pairs that involve these bones.
Note that by line 4, the source chain's lowest common ancestor (\textit{src}) has 1-to-1 correspondence with the target chain's lowest common ancestor (\textit{tgt}).
This satisfies the intuition provided in the first paragraph of this section.
If either chain only has one bone, then we can add the pair of chains as a 1-to-many pair.
Else, we can first add as many 1-to-1 correspondence from the pair of chains, then we are left with a 1-to-many correspondence (lines 15-20).
Here is a simple example: if \textit{src\_chain} is [3, 4, 5, 6] and \textit{tgt\_chain} is [2, 8], then we get one 1-to-1 correspondence pair (3, 2) and one 1-to-many correspondence pair ([4, 5, 6], 8).

\subsection{Ablations on Correspondence Heuristics} \label{subsec:ablation}
\begin{table}[t!]
  \caption{Ablations on cost function heuristics. Each number denotes correspondence accuracy on 36 pairs of manually labeled characters with diverse skeleton topologies.
  ``b-to-b" denotes our additional branch-to-branch cost (Equation~\ref{eqn:btb}), ``len", ``pos", ``hier", and ``oct" are the first, second, third, and fourth term of Equation~\ref{eqn:ltl}.
  The first column shows different combinations of these four terms. 
  }
  \label{tab:ablation}
  \begin{tabular}{c|cc}
    \toprule
    Accuracy & \textbf{With b-to-b} & Without b-to-b \\
    \midrule
    len & 0 & 0\\
    pos & 0.06 & 0.03\\
    hier & 0.03 & 0.03\\
    oct & 0.19 & 0\\
    len, pos & 0.06 & 0.03\\
    hier, oct & 0.42 & 0.13\\
    len, pos, hier & 0.17 & 0.05\\
    pos, hier, oct & 0.64 & 0.19\\
    \textbf{len, pos, hier, oct} & \textbf{0.86} & 0.17\\
  \bottomrule
\end{tabular}
\end{table}

Table~\ref{tab:ablation} shows ablations on the heuristics detailed in the \textit{Leaf-to-leaf} and \textit{Branch-to-branch} paragraphs of Section~\ref{subsec:heuristics}.
Correspondence accuracy is evaluated on 36 pairs of artist labeled characters with diverse skeleton topologies.
Note that ``len, pos" solely uses difference in bone geometry as cost, and ``hier, oct" solely uses difference in bone semantics as cost. 
We have found that combining all bone geometry and semantics heuristics, with the addition of our branch-to-branch cost, drastically increases performance and achieves the highest correspondence accuracy (31 out of 36 pairs).
See Fig.~\ref{fig:moreCorrComp} for some correspondence examples and failure cases.

\section{Unified Skeleton Generation}

\begin{algorithm}[t!]
\SetAlgoLined
\LinesNumbered
\caption{Generating a Unified Skeleton}
\label{algo:uniSkelGen}
\KwIn{Source and target skeletons $\mathbb{S}, \mathbb{D}$, correspondences, step $t$}
\KwOut{Unified skeleton $\mathbb{K}$}
pairs $\leftarrow$ [$(S_0, D_0), (S_1, D_1), ...$], $S \subset \mathbb{S}, D \subset \mathbb{D}$ \\
$\mathbb{K} = []$
\For{$(S, D)$ in pairs}{
    \If{1-to-1}{
        $s \leftarrow S$[0], $d \leftarrow D$[0] \tcp{s, d are corresponded 1-to-1}
        $k \leftarrow$ \textbf{constrained} unified bone.\\
        Set $R_k = (s, d)$ \\
        $\mathbf{h}_k = \text{lerp}(\mathbf{h}_s, \mathbf{h}_d, t)$,
        $l_k = \text{lerp}(l_s, l_d, t)$ \\
        $\mathbf{b}_k = \text{lerp}_{\text{box}}(\mathbf{b}_s, \mathbf{b}_d, t)$,
        $M_k = \text{slerp}(M_s, M_d, t)$\\
        \tcp{*.par denotes the parent of *}
        $k.\text{par} \leftarrow k'$ in $\mathbb{K}$ with $R_{k'} = (s.\text{par}, d.\text{par})$ \\
        Add $k$ to $\mathbb{K}$.
    }\ElseIf{1-to-many}{
        $\{s_i\} \leftarrow S, d \leftarrow D$ \tcp*{assume len(S)>1, len(D)=1}
        $\{d'_i\} \leftarrow$ split $d$ s.t. $\{l_{d'_i}\}$ are proportional to $\{l_{s_i}\}$ \\
        $\{k_i\} \leftarrow$ \textbf{loose} unified bones.\\
        tmp\_bones = [] \\
        \For{$(s_i, d'_i)$ in pairs}{
            Set $R_{k_i} = (s_i, d'_i)$ \\
            $\mathbf{h}_{k_i} = \text{lerp}(\mathbf{h}_{s_i}, \mathbf{h}_d, t)$,
            $l_{k_i} = \text{lerp}(l_{s_i}, l_d, t)$ \\
            $\mathbf{b}_{k_i} = \text{lerp}_{\text{box}}(\mathbf{b}_{s_i}, \mathbf{b}_d, t)$,
            $M_{k_i} = \text{slerp}(M_{s_i}, M_d, t)$\\
            \textbf{if} $i = 0$ \textbf{then} $k_i.\text{par} \leftarrow k'$ in $\mathbb{K}$ with $R_{k'} = (s_0.\text{par}, d.\text{par})$ \\
            \textbf{else} $k_i.\text{par} \leftarrow$ tmp\_bones[$i - 1$] \\
            Add $k_i$ to tmp\_bones.
        }
        Add tmp\_bones to $\mathbb{K}$.
    }\ElseIf{1-to-void}{
        $s \leftarrow S$ \tcp*{assume len(S)=1, len(D)=0}
        $k \leftarrow$ \textbf{virtual} unified bone.\\
        $\mathbf{b}_\mathbb{S}, \mathbf{b}_\mathbb{D} \leftarrow$ bounding boxes of source and target characters, centered at root bones. \\
        $\mathbf{h}_{d'} \leftarrow \mathbf{h}_s$ mapped from $\mathbf{b}_\mathbb{S}$ to $\mathbf{b}_\mathbb{D}$. \\
        $\text{tail}_{d'} \leftarrow \text{tail}_s$ mapped from $\mathbf{b}_\mathbb{S}$ to $\mathbf{b}_\mathbb{D}$. \\
        tmp\_bones = [] \\
        Set $R_k = (s, \emptyset)$ \\
        $\mathbf{h}_k = \text{lerp}(\mathbf{h}_s, \mathbf{h}_{d'}, t)$,
        $l_k = \text{lerp}(l_s, 0, t)$ \\
        $\mathbf{b}_k = \text{lerp}_{\text{box}}(\mathbf{b}_s, \mathbf{b}_0, t)$,
        compute $M_k$ from $\mathbf{h}_{d'}, \text{tail}_{d'}$ \\
        $k.\text{par} \leftarrow k'$ in $\mathbb{K}$ with $R_{k'} = (s.\text{par}, \cdot)$ \\
        Add $k$ to $\mathbb{K}$.
    }
}
\end{algorithm}

\methodname{} preservers a rig throughout interpolation such that intermediate characters are posable, thus enabling interpolation during animation. 
Even though the geometry of the unified skeleton varies depending on step $t$, its topology remains the same. 
Thus, animators can interact with a single unified skeleton and interpolate characters during a motion sequence.
Given input skeletons, correspondence pairs, and an interpolation time step $t$, \methodname{} creates a unified skeleton. 
Algorithm~\ref{algo:uniSkelGen} shows pseudocode for our algorithm that generates unified skeletons.

Algorithm~\ref{algo:uniSkelGen} has different strategies to create unified bones depending on the type of correspondence. 
As detailed in the main paper, 1-to-1 correspondence produces constrained unified bones, 1-to-many correspondence produces loose unified bones, and 1-to-void correspondence produces virtual unified bones.
As a reminder, $R$ denotes the pair of source and target bones that each unified bone references, $\mathbf{h}_k$ denotes the position of the head of a bone $k$, $l_k$ denotes the length of $k$, $\mathbf{b}$ denotes the bounding box of the local surface patch associated with $k$, and $M$ is a matrix composed of the local axes of $k$ where the y-axis is aligned with the direction of the bone.
Please see Section 4 and 5 of the main paper for detailed definitions.

In Algorithm~\ref{algo:uniSkelGen}, ``par" is a shorthand for ``parent".
Lines 31 - 33 involve constructing bounding boxes and mapping between bounding boxes, the details of which can be found in Section 7.1 of the main paper.
In line 37, $\mathbf{b}_0$ is a bounding box whose x, y, z axes lengths are 0.
Line 38 parents $k$ with a unified bone $k'$ who references the parent of $s$ and a target bone.

\begin{figure*}[t!]
  \centering
  \includegraphics[width=0.9\textwidth]{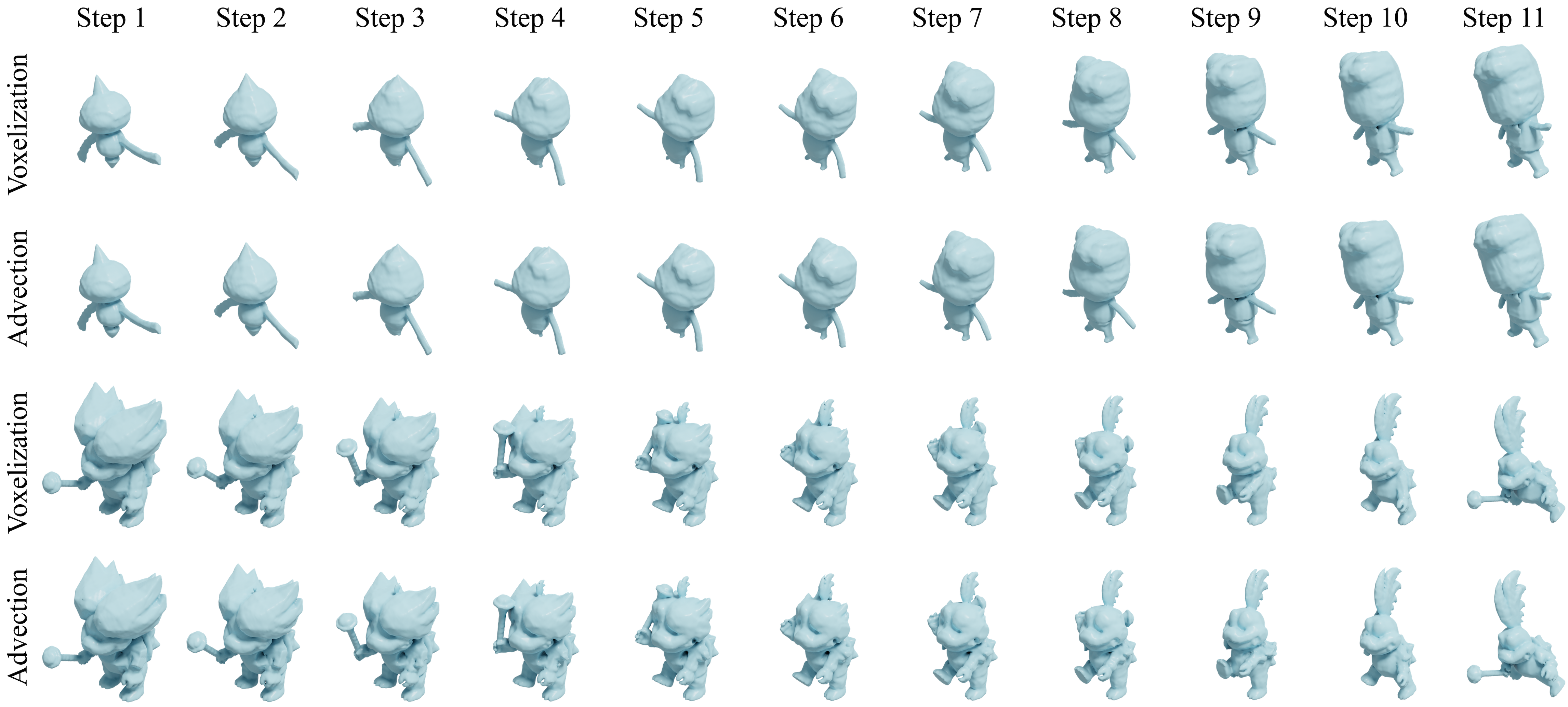}
  \caption{
Qualitative comparison between the slower voxelization method and faster advection method.
The advection method shows no difference for the first character pair and minimal artifacts around the left arm of the second pair.
}
  \label{fig:rbf}
\end{figure*}

\begin{figure*}[t!]
  \centering
  \includegraphics[width=0.7\textwidth]{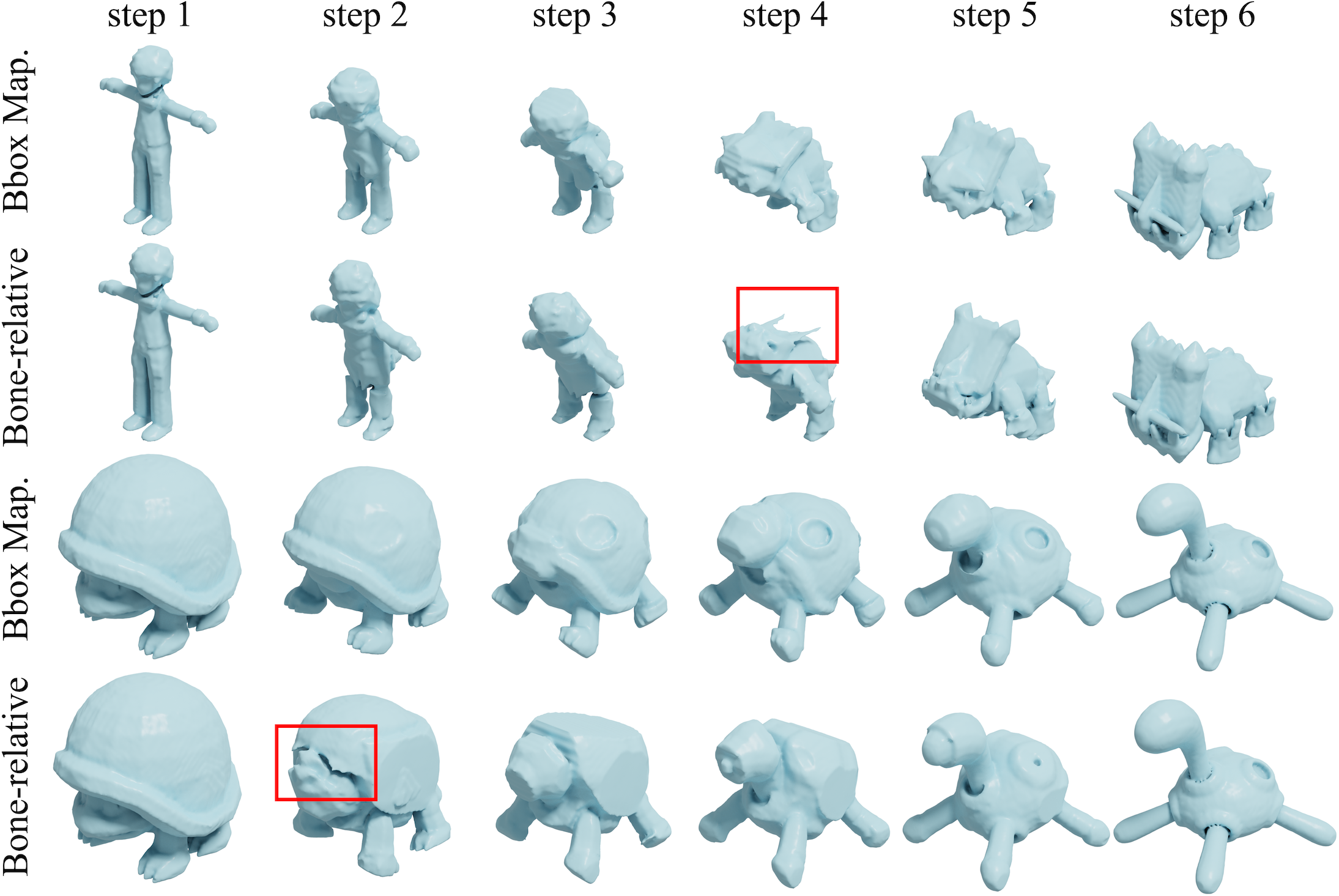}
  \caption{
Qualitative comparison between bounding box mapping (Bbox Map.) and bone-relative mapping.
We observe that bounding box mapping results in smooth interpolation (features of the rhino head gradually appearing in the top row, the circle gradually forming on the shell in the bottom row) while bone-relative mapping causes severe artifacts (red boxes).
}
  \label{fig:bbox}
\end{figure*}

\section{Fast Querying Deformed SDF}

With the unified skeleton's pose retargeted to the source and target skeletons, the two input characters are posed using their skinning weights.
To acquire locally deformed SDF values, the main paper discusses segmenting the posed source and target meshes before turning the segmented parts to SDFs via voxelization and interpolating them. 
The bottleneck lies in voxelization, and we can avoid voxelizing for every pose by taking advantage of the SDFs in rest-pose and the displaced vertices of posed characters. 
Note that the SDFs of characters in rest-pose are pre-computed before the "Skeleton Correspondence" step in Fig. 2 of the main paper.
Given a posed source or target model, \methodname{} builds part-based vector fields using the displacements between locally deformed vertices $V'$ and their rest-pose positions $V$, and it advects query points $p$ in local space by $p' = p + RBF(V' - V, p)$, where $RBF$ queries the vector at $p$ using radial basis interpolation on the sparse data $V' - V$.
Thus, given a query point in a unified bone's space, \methodname{} uses advection to query values from the rest-pose SDFs of source and target bones. 
The advection method runs 17\% faster on average than voxelizing source and target parts for every pose. 
In Fig.~\ref{fig:rbf}, we show comparisons between the two methods and observe that the fast advection method has negligible cost in quality.

\section{More Results}
In this section, we first show ablation on \textit{bounding box mapping} then present more of our skeleton correspondence results and interpolation comparisons with ConvWasser~\cite{ConvWasser} and NeuroMorph~\cite{NeuroMorph}. 
Please watch our supplementary video for animated comparisons and interpolation during animation results. 

\subsection{Ablation on Bounding Box Mapping}
In the main paper, we present a 2D visual comparison between interpolation using bounding box mapping and naive bone-relative mapping that shows the former method preserves characteristics of input source and target geometries.
Fig.~\ref{fig:bbox} shows that CharacterMixer's bounding box mapping component is crucial for obtaining realistic interpolated results because it is feature-preserving.

\subsection{Correspondences}
\begin{figure*}[t!]
  \centering
  \includegraphics[width=0.85\textwidth]{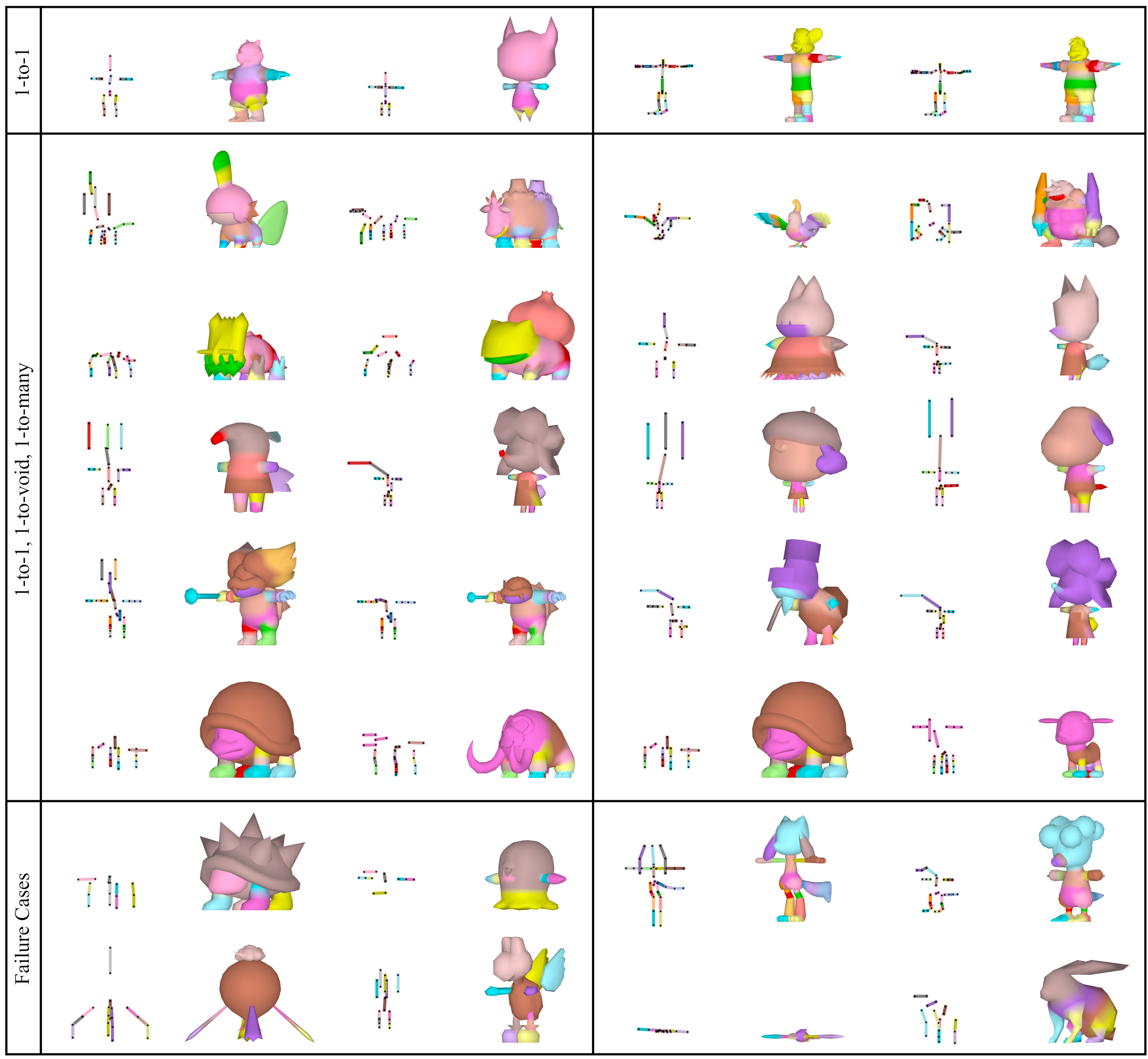}
  \caption{
Correspondence results.
\methodname{} correctly corresponds skeletons of different topologies.
The first row shows examples of only 1-to-1 bone correspondence, and the next five rows show all three types of bone correspondence: 1-to-1, 1-to-void, and 1-to-many.
In the failure cases, each pair only has one or two incorrectly corresponded bones.
For example, for the pair at column 1, row 1, the bottom part of the second character should correspond to void instead of the back left leg of the first character.
}
  \label{fig:moreCorrComp}
\end{figure*}

\begin{figure*}[t!]
  \centering
  \includegraphics[width=0.85\textwidth]{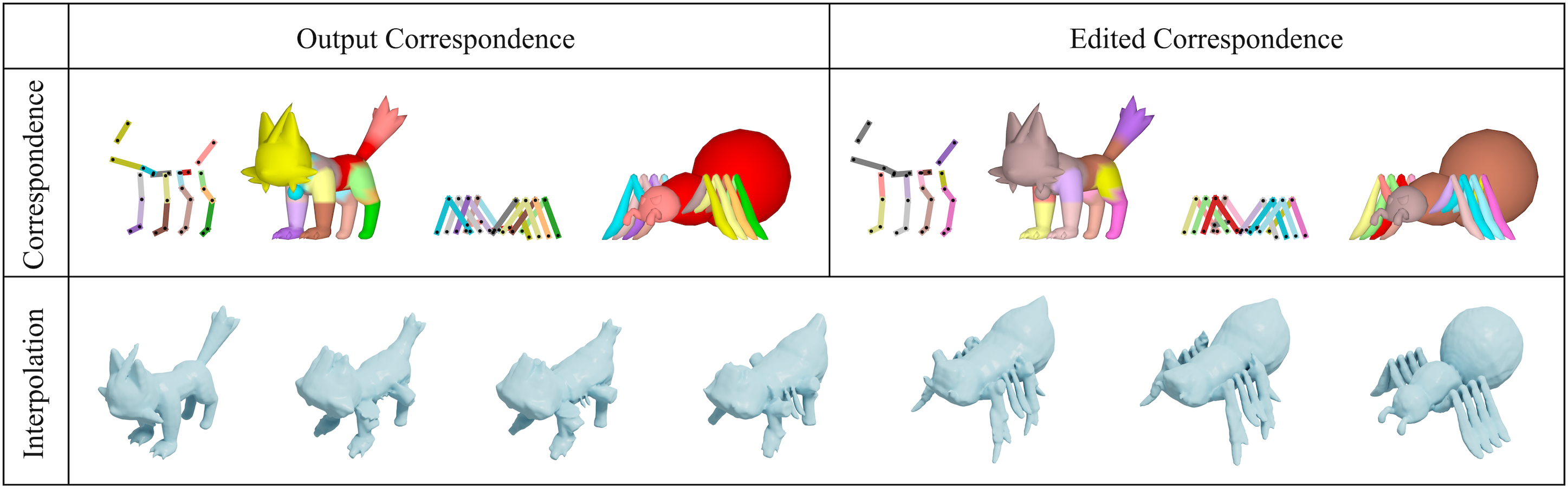}
  \caption{
Extreme case of correspondence and interpolation.
Top left shows output correspondence. 
Top right shows correspondence corrected by an artist. 
Bottom row shows interpolation given the edited correspondence.
}
  \label{fig:catspider}
\end{figure*}

Fig.~\ref{fig:moreCorrComp} shows more correspondences produced by \methodname{}.
For each character pair, we display skeleton correspondences and also visualize them by segmenting each mesh using skinning weights and assigning different colors to surface patches associated with corresponding bones.
Note that for characters in some of the failure cases, it is difficult for artists to determine ``ground truth" correspondences as the skeletons are drastically different. 
Fig.~\ref{fig:catspider} shows an extreme case of correspondence and interpolation between a four-legged cat and an eight-legged spider. 
Even though the output correspondence of \methodname{} is incorrect, \methodname{} allows users to edit and manually define correspondence.
Given edited correspondence, \methodname{} is able to interpolate between the extreme pair of characters.

\subsection{Interpolation Comparisons}

\begin{figure*}[t!]
  \centering
  \includegraphics[width=0.76\textwidth]{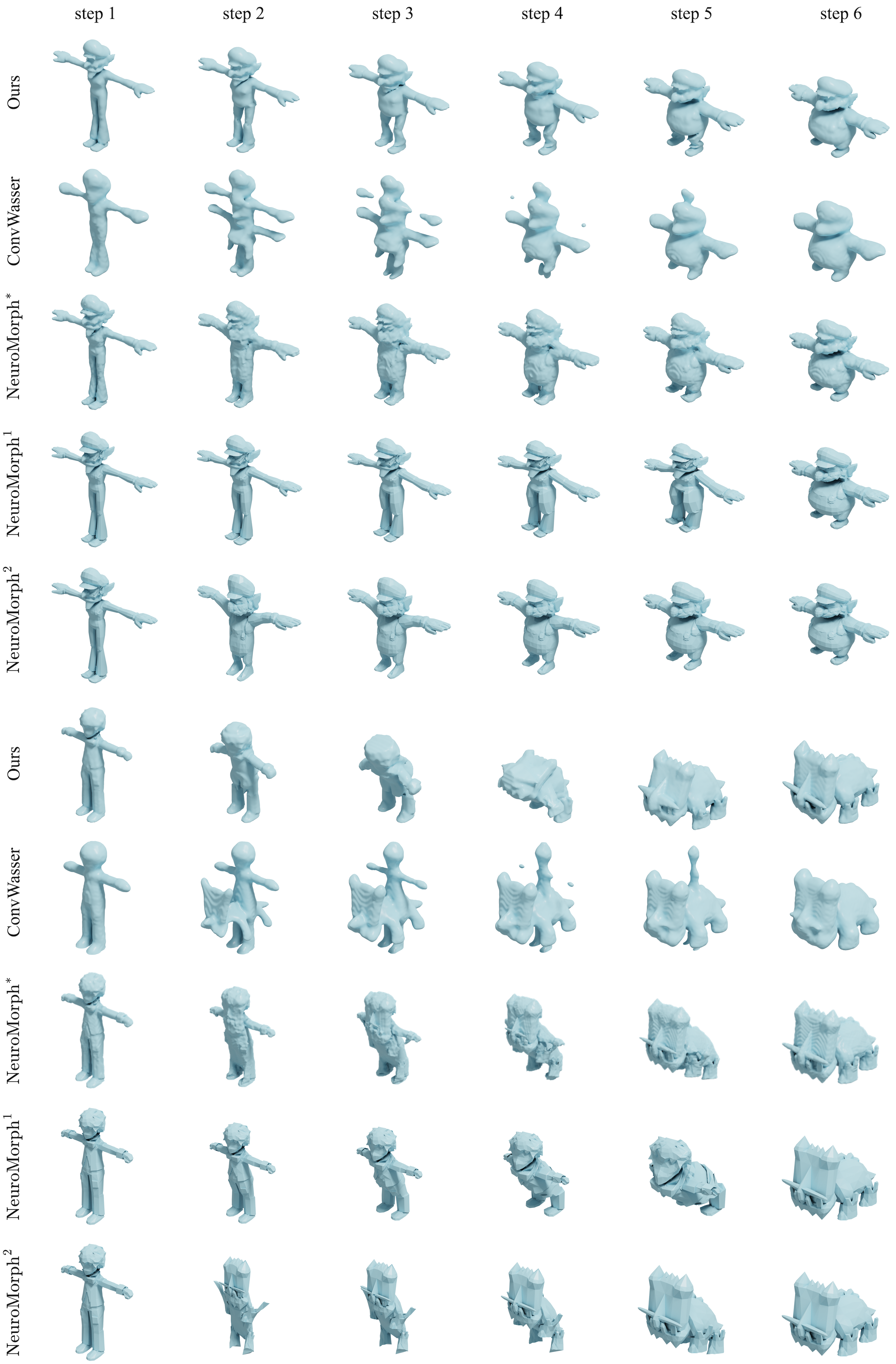}
  \caption{
Qualitative interpolation comparison between our method, ConvWasser~\cite{ConvWasser}, and NeuroMorph~\cite{NeuroMorph}.
Our method produces higher-quality interpolation results.
``$\text{NeuroMorph}^*$" is produced by using NeuroMorph~\cite{NeuroMorph} to interpolate from source to target and from target to source then globally blending the pair of shapes at each time step.
``$\text{NeuroMorph}^1$" shows interpolation from source to target, and ``$\text{NeuroMorph}^2$" shows interpolation from target to source.
Correspondence comparison of the two character pairs can be found in Fig.~\ref{fig:CorrComp_supp}, columns 1 and 2.
}
  \label{fig:20972091_13266516}
\end{figure*}

\begin{figure*}[t!]
  \centering
  \includegraphics[width=0.76\textwidth]{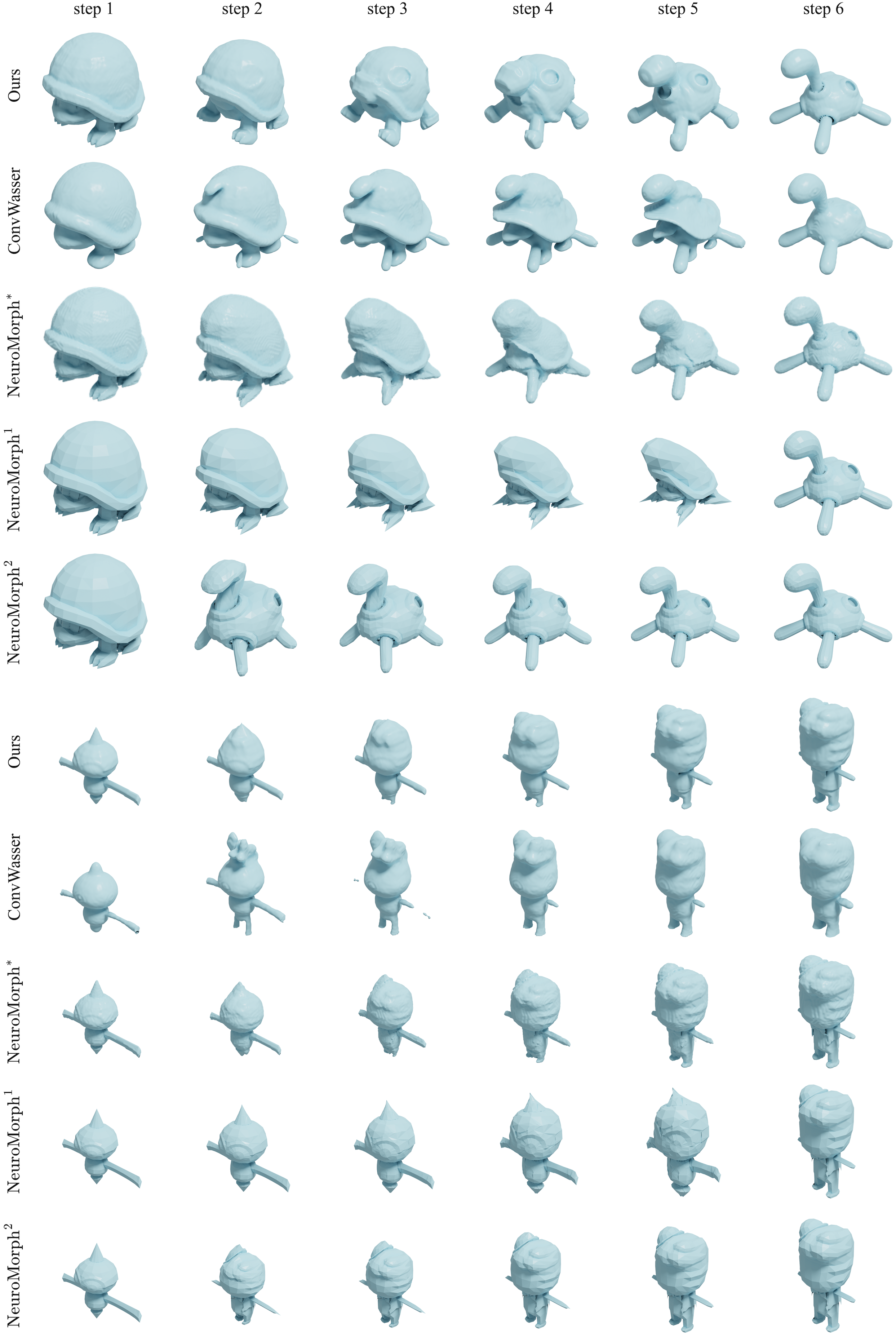}
  \caption{
Qualitative interpolation comparison between our method, ConvWasser~\cite{ConvWasser}, and NeuroMorph~\cite{NeuroMorph}. 
Our method produces higher-quality interpolation results.
``$\text{NeuroMorph}^*$" is produced by using NeuroMorph~\cite{NeuroMorph} to interpolate from source to target and from target to source then globally blending the pair of shapes at each time step.
``$\text{NeuroMorph}^1$" shows interpolation from source to target, and ``$\text{NeuroMorph}^2$" shows interpolation from target to source.
Correspondence comparison of the two character pairs can be found in Fig.~\ref{fig:CorrComp_supp}, columns 3 and 4.
}
  \label{fig:40101919_4321299}
\end{figure*}

In Fig. 7 and 9 of the main paper, we show qualitative comparison of interpolation using our method, ConvWasser~\cite{ConvWasser}, and ``$\text{NeuroMorph}^*$", where ``$\text{NeuroMorph}^*$" is produced by using NeuroMorph~\cite{NeuroMorph} to interpolate from source to target and from target to source then globally blending the pair of shapes at each time step (Section 8 of the main paper).
In Fig.~\ref{fig:20972091_13266516}, and ~\ref{fig:40101919_4321299}, we also show both directions of NeuroMorph's interpolation in addition to ``$\text{NeuroMorph}^*$", where interpolation from source to target is labeled ``$\text{NeuroMorph}^1$", and interpolation from target to source is labeled ``$\text{NeuroMorph}^2$".

\bibliographystyle{eg-alpha} 
\bibliography{main}